%% file: mssm_higgs.tex
\documentclass[onecolumn,amsmath,amssymb,11pt,nofootinbib]{revtex4}

\def\lsim{\raise0.3ex\hbox{$\;<$\kern-0.75em\raise-1.1ex\hbox{$\sim\;$}}}
\def\gsim{\raise0.3ex\hbox{$\;>$\kern-0.75em\raise-1.1ex\hbox{$\sim\;$}}}

 \normalsize
\newcommand{\bmat}{\left(\begin{array}}
\newcommand{\emat}{\end{array}\right)}
\newcommand{\be}{\begin{equation}}
\newcommand{\ee}{\end{equation}}
\newcommand{\bea}{\begin{eqnarray}}
\newcommand{\eea}{\end{eqnarray}}

\usepackage{graphics}
\usepackage{epsfig}

\usepackage{color}
\definecolor{brickred}{RGB}{150,25,14}
\begin{document}
\title{Light sfermion interplay in the 125 GeV MSSM Higgs production and decay at the LHC}
\author{A. Belyaev$^{1,2}$, S. Khalil$^{1,3,4}$, S. Moretti$^{1,2}$, M. C. Thomas$^{1}$ }
\vspace*{0.2cm} \affiliation{ $^1$School of Physics and
Astronomy,
University of Southampton, Highfield, Southampton SO17 1BJ, UK.\\
$^2$Particle Physics Department, Rutherford Appleton Laboratory,
Chilton, Didcot, Oxon OX11 0QX, UK.\\
$^3$Center for Theoretical Physics, Zewail City for Science and Technology, 6 October City, Giza, Egypt.\\
$^4$Department of Mathematics, Faculty of Science, Ain Shams
University, Cairo, Egypt. }
\date{\today}

\begin{abstract}
\noindent\footnotesize
We study the effects from light sfermions on the lightest Higgs boson production and 
decay at the Large Hadron Collider (LHC) within the Minimal Supersymmetric Standard Model (MSSM). 
We find that the scenario with light coloured sfermions -- stops or sbottoms -- 
has the potential to explain a non-universal alteration, as hinted by LHC data,
of the gluon-gluon Fusion ($\mu_{ggF}$) with respect to the Vector Boson Fusion ($\mu_{VBF}$) event rates
and, in particular,  can predict  $\frac{\mu_{VBF}}{\mu_{ggF}}>1$ for all Higgs boson decay channels
in large areas of the parameter space.

We also find that the scenario with a light stop/sbottom
can be complemented by the scenario
in which the total Higgs width, $\Gamma_{\rm tot}$, is reduced due to a suppressed Yukawa coupling $Y_b$.
In this case, the reduction of the Higgs  production rates in the $ggF$ process
which occurs in the maximal mixing scenario is compensated by the reduction of the $H\to b\bar{b}$ partial decay width,
the largest component of $\Gamma_{\rm tot}$.

Furthermore, we highlight the fact that, in the light stop/sbottom scenario,
event rates with the Higgs boson decaying to a $b\bar{b}$ final state are predicted to be essentially below
unity,  especially in case of $ggF$,  which is doubly suppressed, at production,
due to the negative interference from stop/sbottom loops, as well at decay level, due to the $Y_b$  suppression. 
Therefore,  during the future LHC runs, the measurement of $h\to b\bar{b}$ final states 
is a matter of special importance, which will offer additional handles to pin down the possible MSSM
structure of the Higgs sector.

Amongst all viable MSSM configurations that we study (including revisiting a light stau solution), 
we emphasise most the scenario with a light stop, as the latter
is also motivated by Dark Matter and Electro-Weak baryogenesis.
We also perform fits of the MSSM against the LHC data for all scenarios which we introduce,
emphasising the fact that in most cases these are better than for the SM.

\end{abstract}
\maketitle
\newpage
\tableofcontents
\input{01_intro}

\input{02_mssm_para_space}
\input{03.1_mssm_effects}

\input{03.2_stop}
\input{03.3_sbottom}

\input{03.4_stau}

\input{03.5_comb}

\input{04_conclusions}

\section*{Acknowledgments}
We thank Carlos Wagner, Pasquale Di Bari and Ben O'Leary for useful discussions, as well as  Matt Brown for the python code to 
calculate the $\chi^2$ values.
SK thanks The Leverhulme Trust (London, UK) for financial support in the form of a Visiting Professorship to
the University of Southampton.
AB and SM are supported in part by the NExT Institute and MCT is supported by an STFC studentship grant.
We also acknowledge the use of IRIDIS HPC Facility at the University of Southampton for this study.

\bibliographystyle{h-physrev5}
\bibliography{bib}
\end{document}

%% file: 01_intro.tex
\section{Introduction}
The 4$^{\rm th}$ of July 2012 was an important date for the particle physics community, when the
discovery of a Higgs boson with a mass of 125 GeV was announced
by the ATLAS and CMS collaborations~\cite{Chatrchyan:2012ufa,Aad:2012tfa}.
This event was dramatic since a Higgs boson was the last undiscovered particle
desperately searched for to complete the experimental verification of the Standard Model (SM). 
At the same time, the detection of this new state has opened a new chapter in the
exploration of Beyond the SM (BSM) physics, since
many BSM models are consistent with the properties of the discovered Higgs boson
within the accuracy of the experimental data (some are even more preferred by data in comparison to the SM). 
Furthermore, there still remains the need to surpass the
SM from the theoretical side, as the discovered object does nothing to remedy its fundamental flaws: the hierarchy, naturalness and/or fine-tuning problems, the absence of gauge coupling unification at any scale, etc.
Also, the SM does not address fundamental experimental problems on the cosmological scale, such as Dark Matter (DM) /Dark Energy and Electro-Weak (EW) 
Baryo-Genesis (EWBG).

The  recent post-Moriond analysis of  Higgs boson properties
reported by ATLAS~\cite{ATLAS:2013sla} and CMS~\cite{CMS:yva} are
based on 4.7 fb$^{-1}$ at 7 TeV and 13 - 20.7 fb$^{-1}$ at 8 TeV of data
(ATLAS) and  5.1 fb$^{-1}$ at 7 TeV and 19.6 fb$^{-1}$ at 8 TeV of
data (CMS). The results are presented for various Higgs boson
production and decay  channels. The production modes include
gluon-gluon Fusion ($ggF$), Vector Boson Fusion ($VBF$),
Higgs-strahlung ($VH$) and associated production with top-quarks
($ttH$) while the studied  decay modes include $h \to \gamma
\gamma$, $ZZ$, $WW$, $\tau^+ \tau^-$ and 
$b\bar{b}$\footnote{Sensitivity to the $h\to Z\gamma$ mode is much less in
comparison, though some limits already exist~\cite{:2013tq}. Similarly,
for Higgs boson invisible decays \cite{Collaboration:2013xy}.}.

The magnitude of the
signal is usually expressed via the ``signal strength'' parameters $\mu$, defined for either the
entire combination of or the individual decay/production modes,
relative to the SM.
In our study
we define individual  $\mu_{XY}$ for a given production ($X$) and decay ($Y$)
channel, in terms of production cross sections $\sigma$ and decays widths $\Gamma$ (in preference to Branching 
Ratios (BRs)):

\begin{equation}
\mu_{X,Y} =
\frac{\sigma^{\rm MSSM}_X}{\sigma^{\rm SM}_X}
\times
\frac{\rm BR^{\rm MSSM}_Y}{\rm BR^{\rm SM}_Y}
=\kappa_X \times \frac{\Gamma^{\rm MSSM}_Y/\Gamma^{\rm MSSM}_{\rm tot}}{\Gamma^{\rm SM}_Y/\Gamma^{\rm SM}_{\rm tot}}
=
\kappa_X \times
\frac{\Gamma^{\rm MSSM}_Y}{\Gamma^{\rm SM}_Y}
\times
\frac{\Gamma^{\rm SM}_{\rm tot}}{\Gamma^{\rm MSSM}_{\rm tot}}
=\kappa_X \times \kappa_Y \times \kappa_h^{-1},
\end{equation}
where, generally, $X=ggF,VBF,VH,ttH$  and $Y=$~$\gamma\gamma$, $WW$, $ZZ$, $b\bar{b}$, $\tau\bar{\tau}$, etc. 
Notice that, in the above equations,
$\kappa_X$ and $\kappa_Y$ are  equal to the respective ratios of the couplings squared while
$\kappa_h$ is the ratio of the total Higgs boson width in the MSSM relative to the SM.
For example, for $ gg\to h \to \gamma\gamma$, we have
\begin{equation}
\mu_{X,Y} \equiv\mu_{ggF,\gamma\gamma} = \kappa_{ggF} \times \kappa_{\gamma\gamma} \times \kappa_h^{-1}
=\frac{\sigma_{ggF^{\rm MSSM}}}{\sigma_{ggF^{\rm SM}}}\times \frac{\Gamma_{h^{\rm MSSM} \gamma\gamma}}{\Gamma_{h^{\rm SM} \gamma\gamma}}\times \kappa_h^{-1}.
\end{equation}
Notice that the combination of individual production and decay channels
which has been done by experimental papers is a non-trivial procedure
which takes into account the efficiency of the
various channels determining in turn the corresponding weights in the overall combination.

The respective results as reported by ATLAS are given
by~\cite{ATLAS:2013sla}%
\bea
\mu(h \to \gamma \gamma) &=& 1.6 \pm 0.3 \\
\mu(h \to Z Z^{(*)})  &=& 1.5 \pm 0.4 \\
\mu(h \to W W^{(*)})  &=& 1.0 \pm 0.3\\
\mu(h \to b \bar{b})  &=& -0.4 \pm 1.0 \\
\mu(h \to \tau \bar{\tau}) &=& 0.8 \pm 0.7%
\label{ATLAS}
\eea %
while from the CMS collaboration one has~\cite{CMS:yva}
\bea%
\mu(h \to \gamma \gamma) &=& 0.77 \pm 0.27 \\
\mu(h \to Z Z) &=& 0.92\pm {0.28} \\
\mu(h \to W W) &=& 0.68 \pm 0.20\\
\mu(h \to b \bar{b}) &=& 1.15 \pm 0.62 \\
\mu(h \to \tau \bar{\tau}) &=& 1.10 \pm 0.41. %
\label{CMS}
\eea%

It is not possible to perform this combination accurately in this phenomenological 
study as for this one needs to know all details on 
various experimental efficiencies for all production and decay channels
which are not publicly available.
Moreover, the overall signal strength $\mu$
for all production channels combined does not carry valuable information
about possible new physics since in  most BSM scenarios
the main production channels $ggF$ and $VBF$ are non-universally altered 
in comparison to the SM.

\begin{figure}[htbp]
\epsfig{file=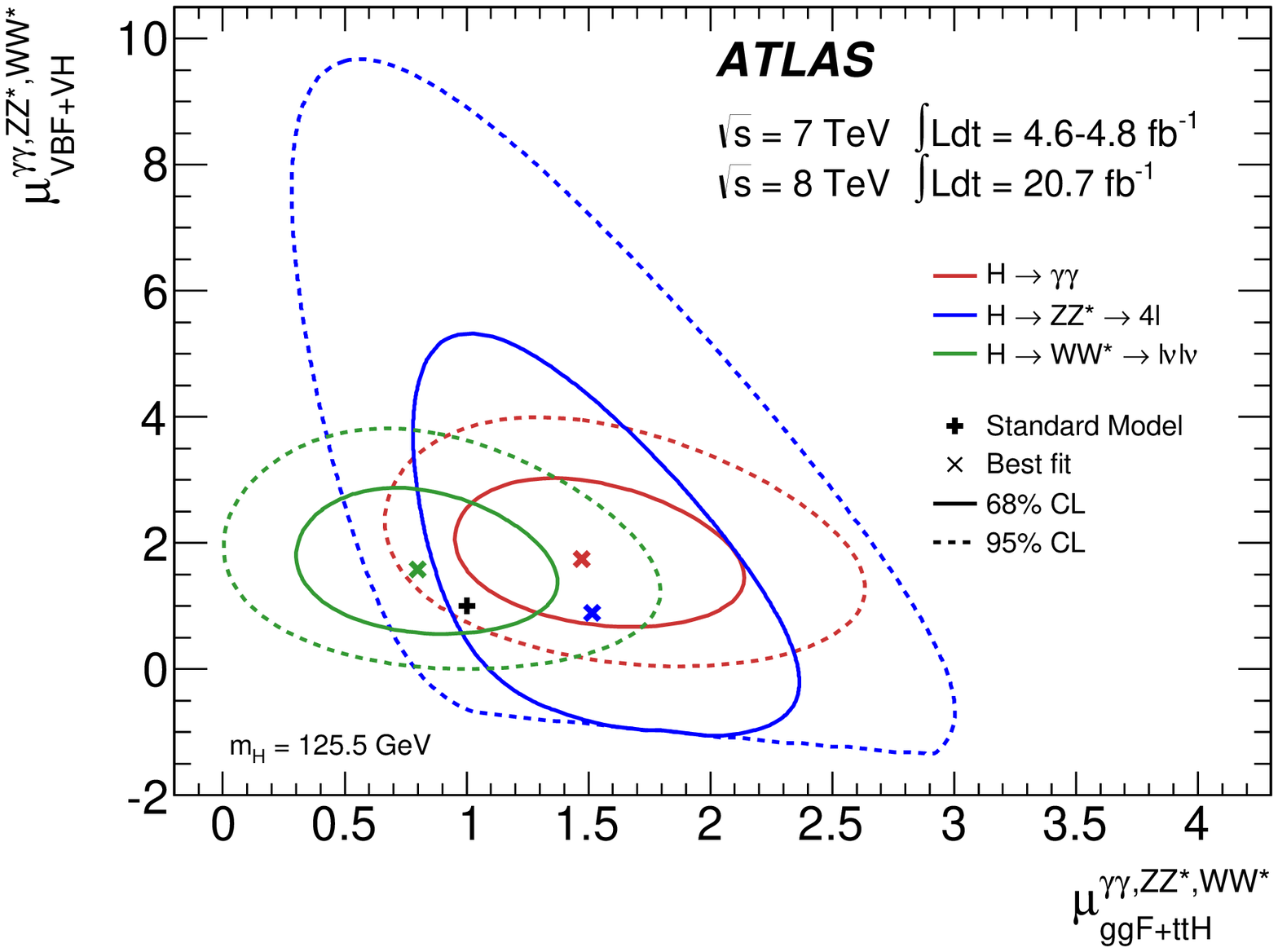,height=6.5cm,width=8.0cm,angle=0}(a)%
\epsfig{file=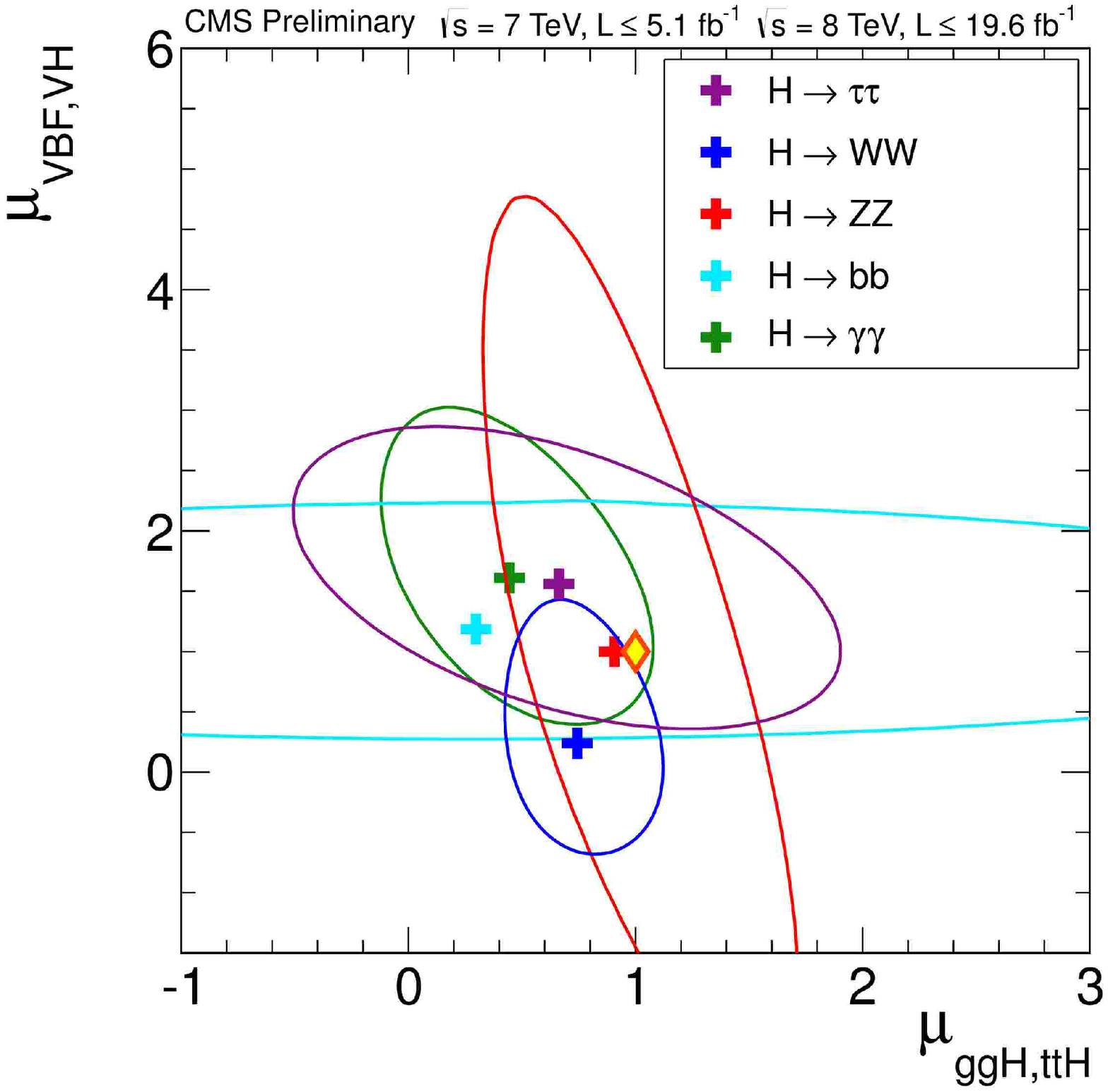,height=6.5cm,width=8.0cm,angle=0}(b)%
\caption{Likelihood contours and best fit values in the ($\mu_{VBF+VH},\mu_{ggF+ttH}$) plane
for different decay channels observed at the LHC:
(a) ATLAS results \cite{ATLAS:2013sla} with 68\% (solid lines) and 95\% (dashed lines)
CL contours and SM expectation (+ symbol);
(b) CMS results \cite{CMS:yva} with 68\% (solid line) CL contours and SM expected value ($\diamond$ symbol). (Herein, the label $ggH$ corresponds
to our $ggF$.)} %
\label{fig:lhc-comb}
\end{figure}

Luckily, both experiments have produced results for the $\mu_{X,Y}$ parameters
for $ggF$ and $VBF$ separately,
as presented in Fig.~\ref{fig:lhc-comb}. Herein, such
results for both collaborations are visualised 
as  likelihood contours in terms of Confidence Level (CL) rates
for the different final states mentioned above.

One can see that these results, on the  one hand, are  consistent with the SM model at 95\% CL
while, on the other hand, there is still a lot of room to accommodate deviations from the SM, 
at least in the  $\pm 40\%$ range at 95\%CL.
One should also notice that for the $h\to \gamma\gamma$  measurement, the ATLAS result
is  about   $2\sigma$ above  the SM prediction for both $ggF$ and $VBF$ production processes,
while the CMS result is approximately $1\sigma$ below the SM value
for  $ggF$ and about  $1\sigma$ above the SM for $VBF$,  respectively.
Thus, one can also see that there is some tension between the ATLAS and CMS results.
From Fig.~\ref{fig:lhc-comb} one can nonetheless see the interesting general pattern
(still within the 1-2$\sigma$ error interval) that
$\mu_{VBF,\gamma\gamma}$  is actually bigger than $\mu_{ggF,\gamma\gamma}$ for both ATLAS and  CMS,
noting that for the CMS collaboration  $\mu_{ggF,\gamma\gamma}$ is essentially below one\footnote{One should also mention that, initially, both collaborations
had initially observed a generic enhancement in the $h\to \gamma\gamma$ channel while, later on, the
CMS results have shifted towards the SM value or even below. }. This trend has been 
quantified by the ATLAS collaboration, who have produced a best fit value of \cite{Aad:2013wqa}
\small \bea%
\frac{\mu_{VBF}}{\mu_{ggF+ttH}} = 1.4^{+0.4}_{-0.3}(\textnormal{stat})^{+0.6}_{-0.4}(\textnormal{sys})
\eea%
for a combination of the $\gamma\gamma$, $ZZ$ and $WW$ data.

On the basis of the pattern of measured 
$\mu_{X,Y}$, it is clear that BSM solutions to the LHC data ought to be investigated 
thoroughly. Herein, in particular, we discuss the case  of the Minimal Supersymmetric Standard Model (MSSM), assess how genuine Supersymmetry (SUSY) effects can affect the Higgs production or decay dynamics (or indeed both)
and draw a picture of the preferred SUSY parameter space in the light of the Higgs LHC data.

In our analysis, we concentrate on the $VBF$ and $ggF$ productions channels only, 
which are the leading ones, 
and limit the study of the decay signatures to the cases of
$h\to \gamma\gamma, WW, ZZ$ final states, as these are the production and decay
modes with the most accurate experimental results.
We should also remark that we carried out our investigation using
renormalisation-group-improved diagrammatic calculations, including
higher-order logarithmic and threshold corrections, using CPsuperH
\cite{Lee:2003nta,Lee:2007gn} (version 2.3).

Quite apart from the fact that current data shows a tendency for
$\mu_{VBF,YY} > \mu_{ggF,YY}$, the LHC
measurements also point to a rather light Higgs mass. While the possibility that the SM
Higgs state had such a mass would be merely a coincidence (as its
mass is a free parameter), in the MSSM, in contrast, the mass of the lightest Higgs boson with SM-like
behaviour is naturally confined to be $\le 135$ GeV \cite{Djouadi:2005gj, Carena:2000yx} by
SUSY itself, which in essence relates trilinear
Higgs and gauge couplings, so that the former are of the same size
as the latter, in turn implying such a naturally small Higgs mass value.
Therefore, to some extent, the Higgs boson mass which is measured at the LHC
favours the MSSM (or some other low energy
SUSY realisation) over the SM, so that it is of the utmost importance to test
the validity of this SUSY hypothesis against the LHC Higgs
data and to establish the viable parameter space.

As we know, the MSSM Higgs sector consists of five Higgs bosons: two CP-even neutral bosons, $h, H$ (with
masses such that $m_h<m_H$)\footnote{We have deliberately used so far the symbol $h$ to signify both the SM
Higgs state and the lightest MSSM CP-even one, as our MSSM solutions to the Higgs data will only involve the latter amongst the possible neutral Higgs states.}, one CP-odd, $A$, and a pair which is
charged, $H^{\pm}$. At tree level, the mixing between the two CP-even neutral Higgs
bosons is defined by the mixing angle $\alpha$, which is a derived
quantity uniquely determined by two independent parameters
which can be taken
as the mass of any of the five physical states (hereafter we take
$m_h$) and  the ratio between the Vacuum Expectation Values (VEVs)
of the two Higgs doublet fields pertaining to the MSSM, denoted by
$\tan \beta$. However, while performing an analysis in higher orders (or in the presence of loop diagrams 
at lowest order as is the case for the $hgg$, $h\gamma\gamma$ and $h\gamma Z$ effective couplings),
one ought to account for the sparticle sector of the MSSM too, which in turn implies the introduction of additional parameters.

Previous literature has explored the Higgs sector in a variety of 
SUSY scenarios, such as the MSSM \cite{Arbey:2012bp,Bechtle:2012jw,SchmidtHoberg:2012ip,
Heng:2012at,Drees:2012fb,Arbey:2012dq,SchmidtHoberg:2012yy,Carena:2012xa,Carena:2011aa,
Hall:2011aa,Heinemeyer:2011aa,Arbey:2011ab,Draper:2011aa,Chen:2012wz,Guo:2011ab,He:2011gc,
Djouadi:2011aa,Cheung:2011nv,Batell:2011pz,Christensen:2012ei} (also the constrained 
version \cite{
Kadastik:2011aa,Baer:2011ab,Aparicio:2012iw,Ellis:2012aa,Baer:2012uya,Desai:2012qy,Cao:2011sn}), 
Next-to-MSSM \cite{King:2012tr,Gunion:2012gc,Belanger:2012tt,Gunion:2012zd,Ellwanger:2012ke,
Ellwanger:2011aa,Cao:2012fz,Kang:2012sy} and
 (B--L)SSM \cite{Elsayed:2011de,Basso:2012tr,Khalil:2012gs,Khalil:2013in}, including scenarios
with light charginos \cite{Batell:2013bka}, staus \cite{Carena:2011aa,Carena:2012gp} 
and stops \cite{Carena:2011aa}.

In our paper we re-examine the light stop, sbottom and stau scenarios, but also
extend previous research by allowing any combination of MSSM quantum corrections, mixing effects
and/or light MSSM fermions entering loops. In particular, we are the first to discuss
how the MSSM could explain a non-universal alteration in $\mu_{VBF,YY}$ 
versus $\mu_{ggF,YY}$ from their SM values such that $\tfrac{\mu_{VBF}}{\mu_{ggF}} \ne 1$,
and use these to examine the compatibility of the MSSM 
against LHC data. We also examine its ability to produce enhanced (with respect to the SM)
rates in the di-photon channel, such that ${\mu_{VBF}}>1$ and/or ${\mu_{ggF}} > 1$
and explore the effects of deviations entering all other measured Higgs boson couplings to SM particles.

The plan of the paper is as follows. 
In Section II we introduce the general setup and the MSSM parameter 
space that we explore, specific to the Higgs sector.
In Section III(A) we study the possible generic MSSM effects on the relevant dynamics,
namely, onto Higgs production, decay and total width.
In Sections III(B) and III(C) we study the effects of stops and sbottoms,
respectively, where we find that both can give rise to non-universal alterations
in $\mu_{VBF}$ versus $\mu_{ggF}$ as both particles are able to affect the 
$ggF$ fusion rate but not the $VBF$ one.
Section III(D) explores the stau contribution, where we find that it can only 
produce a universal increase in cross section in the di-photon channel, 
irrespectively of the production channel, as it only appears in the $\gamma\gamma$ 
(and $Z\gamma$) decay loops.
In view of the Higgs data potentially indicating a non-universality in the production 
channels compared to SM predictions, in Section III(E) we look at the combined
effects of these scenarios as well as perform a $\chi^2$ fit of the MSSM parameter
space with respect to LHC data. We draw our conclusions in Section IV.


%% file: 02_mssm_para_space.tex
\section{MSSM setup and the parameter space}

The MSSM is essentially a straightforward supersymmetrisation of
the SM with the minimal number of new parameters. It is the most
widely studied potentially realistic SUSY model. Furthermore, while
different assumptions about the SUSY breaking dynamics can be made and these in turn
lead to quite different phenomenological predictions, one can always assume an EW scale configuration and
scan over the SUSY parameters relevant at that energy. 

From this point of view, it becomes important to specify the MSSM spectrum (of masses
and couplings). The particle
content of the MSSM is three generations of (chiral) quark and
lepton superfields, the (vector) superfields necessary to gauge
the $SU(3)_C\times SU(2)_L\times U(1)_Y$ group of the SM and two
(chiral) $SU(2)$ Higgs doublet superfields. The introduction of a
second Higgs doublet, with respect to the SM, is necessary in order to cancel the anomalies
produced by the fermionic members of the first Higgs superfield
and also to give masses to both up- and down-type fermions.

The interactions between Higgs and matter superfields are
described by the superpotential
\begin{equation}
W = Y^E_{ij}L_i E^c_j H_d   + Y^D_{ij} Q_i D^c_jH_d  + Y^U_{ij}
Q_i U^c_j H_u + \mu H_u H_d . \label{superpot}
\end{equation}
Here $Q_L$ contains the $SU(2)$ (s)quark doublets and $U_L^c$ and
$D_L^c$ the corresponding singlets, while the (s)lepton doublets and
singlets reside in $L_L$ and $E_L^c$, respectively. In addition, $H_u$ and
$H_d$ denote Higgs superfields with hypercharge $Y=\pm
\frac{1}{2}$. The MSSM assumes certain soft SUSY breaking terms at a
grand unification scale $M_{\textnormal{GUT}} = 3 \times 10^{16}$ GeV. These soft
SUSY breaking terms are categorised as trilinear scalar couplings
$A^f_{ij}$, gaugino masses $M_a$, sfermion mass-squared terms
$\tilde{m}^f_{ij}$, and bilinear scalar coupling $B$.

In the MSSM, the SM-like Higgs is the lightest CP-even Higgs, which is
defined as
$$h =\sin\alpha~ {\rm Re}(H_d^0) + \cos\alpha~
{\rm Re}(H_u^0), $$ with mixing angle $\alpha$ given by
\be%
\tan 2 \alpha = \tan 2 \beta ~ \frac{M_A^2+M_Z^2}{M_A^2 - M_Z^2}.%
\ee%
The mass of the MSSM Higgs can be written, at one loop, as \cite{Ellis1990,Haber1991,Haber:1996fp} %
\be %
m_h^2 \simeq M_Z^2 \cos^2 2 \beta + \frac{3}{4 \pi^2}
\frac{m_t^4}{v^2} \left[\log\left(\frac{M_S^2}{m_t^2} \right) +
\frac{X_t^2}{M_S^2} \left(1-\frac{X_t}{12 M_S^2}\right)\right],%
\ee %
where $M_S^2 = \frac{1}{2} (M_{\tilde{t}_1}^2 +
M_{\tilde{t}_2}^2)$ and $X_t = A_t - \mu \cot \beta$. From this
expression, one can easily show that the maximum value of $m_h$ is
obtained at the maximal stop mixing, {\it i.e.}, at $X_t =
\sqrt{6} M_S$. Also, in order to have $m_h = 125$ GeV, one should
assume that at least one stop has a mass of ${\cal O}(1)$ TeV, while the
other stop can be light. In addition, a quite large stop mixing is
required, {\it i.e.}, $A_t \gsim 1$ TeV.

In our analysis we are interested in the stop, sbottom and stau
states as light particles. In general, one can write the squared
mass matrices of these particles in the basis of the
gauge eigenstates $(\tilde{f}_L, \tilde{f}_R)$ as%
\be%
M^2_{{\tilde{f}}} = \left( \begin{array}{cc} m^2_f + m^2_{LL} &
 m_{f} X_f \\
m_{f} X_f &  m_f^2 + m^2_{RR} \end{array}
\right),\label{eq:stopmat}
 \ee
where %
\bea%
m^2_{LL} &=& m^2_{\tilde{f}_L} + (T_{3f} - Q_f s_W^2) M_Z^2 \cos
2 \beta , \\
m^2_{RR} &=& m^2_{\tilde{f}_R} +  Q_f s_W^2 M_Z^2 \cos 2 \beta , \\
X_f &=& A_f - \mu (\tan \beta)^{-2 T_{3f}} , %
\eea%
where $T_{3f}$ is the third component of the weak isospin and
$Q_f$ is the electric charge.
Thus, the  sfermion physical masses are given by%
\be%
m^2_{\tilde{f}_{1,2}} = m^2_f + \frac{1}{2} \left[m^2_{LL} +
m^2_{RR} \mp \sqrt{(m_{LL}^2 - m_{RR}^2)^2 + 4 m_f^2
X_f^2}~\right]\, , %
\ee%
and the mixing angles are given by%
\be%
\tan 2 \theta_f = \frac{2 m_f X_f}{m_{LL}^2 -
m_{RR}^2} \, . %
\ee%
In this regard, one notices that the mixing in the stop sector is
very strong, hence one of the stops, $\tilde{t}_1$, can be very
light. Also, with large $\tan \beta$ and $\vert \mu \vert$, the
mixing in the sbottom and stau sectors can also be very strong,
therefore light $\tilde{b}_1$ and $\tilde{\tau}_1$ are further
obtained.

LHC constraints on SUSY masses are generally quoted as around 600--700 GeV
for stops and sbottoms and in the region of 
300 GeV for staus ~\cite{ATLAS:2013sla,CMS:yva}, depending on assumptions
regarding the decay processes and the masses of decay products. 
However,
these results all rely strongly on a sizeable mass splitting between these sparticles
and the Lightest Supersymmetric Sparticle (LSP), a neutralino, to which they decay.
These limits are drastically reduced in the region of low mass splittings: {\it e.g.},
if $m_{\tilde{t}} \approx m_t + m_{\tilde{\chi}^0}$, then the stop 
signal becomes difficult to distinguish from the $t \bar{t}$ background
and the LHC data are unable to constrain the stop mass.
A similar situation arises for other mass splitting scenarios,
such as when the stop mass is close to the mass of the LSP  
$m_{\tilde{t}} \approx m_c + m_{\tilde{\chi}^0}$.
In this case the stop mass limit is reduced
down to the LEP limit $\sim$ 95 GeV~\cite{lep2-sel}.
The limits for sbottom and stau masses can also be markedly reduced down to LEP limits
($\sim$ 95 GeV for sbottoms and $\sim$ 85 GeV for staus~\cite{lep2-sel}) for
appropriate mass splittings.

In the present study, we performed a large scan of parameter space using
CPsuperH to produce the data points, concentrating on those
parameters with an important role in the masses and couplings of
the stops, sbottoms and staus as well as the mass of the Higgs
boson and its couplings to the bottom quark\footnote{Recall, in
fact, that the dominant component of the Higgs boson width for 
masses of order 125 GeV is typically the partial width in $b\bar
b$ pairs.}. These masses and couplings are largely independent 
of the $M_1$ mass parameter (they vary only $\sim$ 0.1\%  for $M_1$ ranged from 0.1 TeV - 100 TeV).
However when $M_1 \ll (M_2, \mu)$, the lightest neutralino mass, $m_{\tilde{\chi}_1^0} \approx M_1$, 
so that for any point in our parameter scan its values can be chosen to
give whichever LSP mass is required to be consistent with
cosmological and LHC constraints, without otherwise altering our conclusions. 
In Tab. \ref{range}, we list the range of parameters
of this scan.

\begin{table}[h]
\begin{tabular}{|c|l|c|l|}
 \hline\hline
  ~~~~Parameter~~~~ 		& ~~~~~~Range~~~~~~ 	& ~~~~Parameter~~~~   & ~~~~~~Range~~~~~~ \\
  \hline
  $\tan \beta$ 			& $[2,50]$ 		& $M_{Q3}$ 		& $[0.1,10]$ TeV \\
  $M_{H^\pm}$ 			& $[0.2,2]$ TeV		& $M_{U3}$		& $[0.1,5]$ TeV  \\
  $\mu$ 			& $[0.1,5]$ TeV         & $M_{D3}$ 		& $[0.1,20]$ TeV \\
  $A_t$ 		        & $[0.1,10]$ TeV        & $M_{L3}$		& $[0.1,5]$ TeV \\
  $A_b$ 			& $[0.1,10]$ TeV        & $M_{E3}$		& $[0.1,5]$ TeV  \\
  $A_\tau$ 			& $[0.1,5]$ TeV         & $M_3$	         	& $[0.1,5]$ TeV \\
  $A_e,A_\mu,A_u,A_d,A_c,A_s$  	& fixed at 10 GeV 	& $M_2$  		& fixed at 3 TeV \\
\hline\hline
\end{tabular}
\caption{Range of scanned parameters. $M_1$ can be chosen to provide an LSP (neutralino) mass to overcome 
cosmological and LHC constraints without altering any other relevant results.}\label{range}
\end{table}

To increase the number of points in the parameter space of interest,
three further localised scans were performed, in each case reducing the scanned 
range of one variable, with the other variable ranges remaining as described in Tab. \ref{range}.
The altered ranges in these additional scans were:
\begin{enumerate}
  \item $100$ GeV $ \leq M_{U3} \leq 300$ GeV to produce light stops;
  \item $100$ GeV $ \leq M_{D3} \leq 400$ GeV to produce light sbottoms;
  \item $100$ GeV $ \leq M_{E3} \leq 400$ GeV and $100$ GeV $ \leq M_{L3} \leq 400$ GeV to produce light staus.
\end{enumerate}

In order to avoid colour breaking minima of the $\tilde{t}$ or $\tilde{b}$ fields, we apply the 
$|A_t|, |\mu| \le 1.5 (M_{Q3} + M_{U3})$
constraints to all plots and numerical results unless otherwise stated \cite{Drees:2005jg,Frere:1983ag,Claudson:1983et}.
These requirements are somewhat  conservative, in the light of a very recent analysis in Ref.~\cite{Camargo-Molina:2013qva},
yet we maintained them in order to simplify our study.


%% file: 03.1_mssm_effects.tex

\section{MSSM effects in Higgs production and decay}
In this Section, we discuss MSSM effects 
which alter the Higgs event rates at the LHC 
as compared to those of the SM. We start with a first Subsection, in turn divided in the three  parts corresponding to introducing the structure of Higgs cross sections, di-photon decay and total width. The remaining four Subsections deal with stop, sbottom, stau and
their combined effects, respectively, in either of these contexts.
 
\subsection{The three contexts for MSSM effects}
\label{sec:MSSM_effects}

\subsubsection{MSSM Higgs production}

We start our discussion with MSSM Higgs boson production via the gluon-gluon fusion
process, which is the dominant channel for Higgs searches at the LHC.
In the SM, this mode is predominantly  mediated
by top quarks via a one-loop triangle diagram
while the contribution from other quarks, even the bottom one,
is only at the few percent level.
 
In the MSSM, however, strongly interacting superpartners of the SM quarks, {\it i.e.}, the
squarks, could provide a sizeable contribution to this triangle loop.

The lowest order parton-level cross section
can be written as
\be%
\hat{\sigma}_{\rm LO}(gg\rightarrow h)= \frac{\pi^2}{8m_h}
\Gamma_{\rm LO} (h \rightarrow gg) \Delta(\hat{s}-m_h^2),%
\ee%
where $\hat{s}$ is the center-of-mass energy at the partonic level and
$\Delta(\hat{s}-m_h^2)$ is the Breit-Wigner form of the Higgs
boson propagator, which is given by
$$\Delta(\hat{s}-m_h^2) = \frac{1}{\pi} \frac{\hat{s}
\Gamma_h/m_h}{(\hat{s}-m_h^2)^2 + (\hat{s} \Gamma_h/m_h)^2},$$
and $\Gamma_h$ is the total Higgs boson decay width, while
its partial decay width, $\Gamma_{\rm LO}(h \to gg)$, 
is given by
\be %
\Gamma_{\rm LO}(h \rightarrow g g) =\frac{\alpha_s^2 m_h^3}{512 \pi^3}
\Big| 
\sum_{f}\frac{2 Y_f}{m_f} F_{1/2}(x_f) 
+
\sum_{S}\frac{g_{hSS}}{m^2_{S}} F_{0}(x_{S})
\Big|^2, %
\label{eq:h-gg}
\ee%
where $Y_f$ and $g_{hSS}$ are the MSSM
Higgs couplings to the respective (s)particle species
for fermion (spin-1/2) and scalar (spin-0)
particles, respectively, entering the triangle diagram.
The loop functions $F_{1/2,0}$ can be
found, for example,  in \cite{Carena:2012xa}. Here, $x_i$ is defined as
$4m_i^2/m_h^2$, with $m_i$ being the mass running in the loop.
In the decoupling (or quasi-decoupling) 
regime, as in the case of the SM limit of the MSSM, the
top quark contribution is dominant among the quarks,
since it has the largest Yukawa coupling, 
while the contribution from the other quarks 
(mainly coming from the bottom quark) is at the percent level, as intimated.
The role of the bottom quark can be dramatically different though  
in the non-decoupling regime, when the $hbb$ Yukawa coupling, 
$Y_b^{\rm MSSM} =
-\frac{m_b}{v}\frac{\sin\alpha}{\cos \beta}=Y_b^{\rm SM}\frac{\sin\alpha}{\cos \beta}$ is
enhanced
by $\sin{\alpha}/\cos\beta\simeq \tan\beta$ in comparison to the SM,
enabling the bottom quark contribution to the triangle loop to increase and
even dominate over the top quark for large values of $\tan\beta$.
However, this is not a realistic possibility, since LHC data do not indicate such significant deviations
of the Higgs couplings from SM the values (they are within  
a 50\% or so range from the latter), while data  
on the 
Higgs mass measurement  indicate
that, if the MSSM  is realised in Nature, then the decoupling or quasi-decoupling regime should take place. In fact,
the Higgs boson mass is close to the one reached in the decoupling limit, requiring $\alpha \approx
 \beta - \frac{\pi}{2}$, hence $Y_b^{\rm MSSM} \approx Y_b^{\rm SM}$ as well as 
 $Y_t^{\rm MSSM} \equiv \frac{m_t}{v}\frac{\cos\alpha}{\sin \beta}\approx Y_t^{\rm SM}$.

From eq. (\ref{eq:h-gg}) one can see that the $g_{hSS}$ coupling has dimension one,
while it is more convenient to define a dimensionless $\hat{g}_{hSS}$ to be used hereafter:
\bea
\hat{g}_{hSS}=\frac{{g}_{hSS}}{M_W/g} = \frac{g_{hSS}}{(4\sqrt{2}G_F)^{-\frac{1}{2}}}= g_{hSS}\sqrt{4\sqrt{2}G_F},
\eea
where $G_F$ is the Fermi constant.
So $\Gamma_{\rm LO}(h \rightarrow g g)$
will have a form
 \be %
\Gamma_{\rm LO}(h \rightarrow g g) =\frac{\alpha_s^2 m_h^3}{512 \pi^3}
\Big| 
\sum_{f}\frac{2 Y_f}{m_f} F_{1/2}(x_f) 
+
\sum_{S}\frac{\hat{g}_{hSS}}{m^2_{S}}\frac{M_W}{g} F_{0}(x_{S})
\Big|^2. %
\label{eq:h-gg-b}
\ee%

One should also note that the functions $F_{1/2}(x)$ and $F_{0}(x)$ reach a plateau very quickly 
for $x>1$ and their values are about 1.4 and 0.4, respectively.
This fact has important consequences, which we will discuss together with the Higgs decay into two photons,
in the next Subsection.
%
%
%
%
The specific effects of stop and sbottom loops will be discussed in 
Sections III(b) and III(c).

\subsubsection{MSSM Higgs decay into di-photons}

In the SM, the one-loop partial decay width of the $h$ state into two photons is
given by \cite{Shifman:1979eb} \cite{Ellis:1975ap}
\bea
\Gamma(h \to \gamma \gamma)&=&\frac{G_F\alpha^2 m_h^3}{128\sqrt{2} \pi^3}\Big
\vert F_1(x_V) +  \sum_{f} N_{c,f} Q_f^2 F_{1/2}(x_f) \Big\vert^2=
\frac{\alpha^2 m_h^3}{1024 \pi^3} \frac{g^2}{M_W^2}\Big
\vert F_1(x_V) +  \sum_{f} N_{c,f} Q_f^2 F_{1/2}(x_f) \Big\vert^2
\nonumber
\\
&=&\frac{\alpha^2 m_h^3}{1024 \pi^3} \Big
\vert \frac{g_{hWW}}{M_W^2}  F_1(x_V) +  \sum_{f}\frac{2
Y_f}{m_f} N_{c,f} Q_f^2 F_{1/2}(x_f) \Big\vert^2
\eea
while in the
MSSM the one-loop partial decay width of the $h$ state into two photons also
gets a contribution from scalar particles represented by sfermions and charged Higgs boson
and is given by
\bea
\Gamma(h \to \gamma \gamma)=\frac{\alpha^2 m_h^3}{1024 \pi^3} \Big
\vert \frac{g_{hWW}}{M_W^2} F_1(x_V) +  
\sum_{f}\frac{2Y_f}{m_f} N_{c,f} Q_f^2 F_{1/2}(x_f) +
\sum_{S}\frac{\hat{g}_{hSS}}{m^2_{S}}\frac{M_W}{g} N_{c,S} Q_S^2 F_0(x_S) \Big\vert^2,
\eea
where  $V, f$, and $S$ stand for Vector, fermion and scalar
particles respectively, entering the one-loop triangle diagram,
$g_{hWW}$ is the MSSM
Higgs coupling to $W$-boson, while  $Y_f$ and $\hat{g}_{hSS}$ are the MSSM
couplings of Higgs boson to fermions and scalars defined in the previous Subsection.

The genuine SUSY contributions to $\Gamma(h\to \gamma\gamma)$ are
mediated by charged Higgs, charginos and charged sfermions. The SM-like
part is dominated by $W$-gauge bosons, for which  $F_1(x_W) \simeq
-8.3$, whereas the top quark loop is subdominant and enters with opposite sign,
$N_{c,f} Q_f^2 F_{1/2}(x_f) \simeq 1.8$, with all other
fermions contributing negligibly. It is also worth mentioning that 
$F_0(x_S)\sim 0.4$,  which is about a factor 20  smaller than $F_1(x_W)$
and approximately a factor 4 smaller than  $F_{1/2}(x_f)$.

Keeping this in mind, let us discuss possible sources of the enhancement  of  the $h \to \gamma
\gamma$ effective coupling  which in the MSSM may come through one of the following
possibilities: $(a)$ by the induction of a large scalar contribution, due to
the light stop or/and sbottom  or/and stau, with negative coupling $\hat{g}_{hSS}$ so
that it interferes constructively with the dominant
$W$-contribution;
$(b)$ via charged Higgs boson contributions;
$(c)$ via chargino contributions;
$(d)$ via modification of the Yukawa couplings of top and bottom quarks in the loop.
In the decoupling or quasi-decoupling regimes which eventually take place,
as discussed above,  scenario $(d)$ does not occur.
As for case $(b)$, then 
taking into account that the charged Higgs mass is 
limited to be above 200 GeV (see e.g. \cite{Christensen:2012ei,Chakraborty:2013si} 
and references there in), the fact that its loop contribution is suppressed by a factor of $(M_W/M_{H^\pm})^2$
and that $\hat{g}_{hH^+H^-}$ is of the  order of the electroweak coupling
(contrary to the $\hat{g}_{hSS}$  coupling for squarks and sleptons which can be large as we discuss below),
we have found that the contribution from charged Higgs bosons is generally negligible.
In  case $(c)$, the chargino contribution can be bigger than that of the charged Higgs,
because of the ratio $F_{1/2}:F_{0} \simeq 4$ and because the chargino has a lower mass limit 
of approximately 100 GeV (coming from LEP2~\cite{Abbiendi:2003sc}).  We have found that the
maximum chargino contribution  is reached in the  $\mu \to M_2,\ \ \tan\beta \to 1 $ limit 
(where $\mu$ is the Higgs mass parameter
while $M_2$ is the gaugino soft breaking mass)
and can enhance the SM $h \to \gamma \gamma$ partial decay width by about 30\%. This agrees with the 
recent results of~\cite{Batell:2013bka}.
The scenario with very light charginos is not the focus of our paper, where we assume charginos 
to have a mass of at least a few hundred 
GeVs, and for which  the virtual chargino contribution to the $h \to \gamma \gamma$ decay is  negligible.
Moreover, the effect from the light charginos which could alter only the $h \to \gamma \gamma$ decay
is qualitatively similar to the effect from the light staus, which quantitatively can be much larger~\cite{Carena:2012gp},
and which we consider in the current study in great detail together with the light sbottom and light stop scenarios.
Therefore, in this study we concentrate on scenario $(a)$ in which sizeable MSSM contributions
via scalar loops are still possible.

It is worth stressing again two
important details related to the
scalar contribution to $h \to \gamma \gamma$.
Firstly, the
smallness of the loop function $F_0$ with respect to $F_{1/2}$
and with respect to $F_1$ too, and, secondly,
the mass suppression factor, $(M_W/M_S)^2$,
mentioned above.
Therefore the only way to have a sizeable effect from the scalar loops
is to be in a scenario with large 
coupling $\hat{g}_{hSS}$ and light scalars.
In such a scenario the scalar loop competing with the fermion loop has a larger relative
contribution to $\Gamma(h \to gg)$ than to $\Gamma(h \to \gamma\gamma)$
where it would also compete with the dominant vector boson loop.
At the same time,
the contribution from squarks is  opposite 
for Higgs production via  gluon-gluon fusion compared to di-photon decay:
depending on the sign of $\hat{g}_{hSS}$, 
they will  destructively (constructively) interfere with top quarks in production loops
and constructively (destructively)  interfere with $W$-boson loops in Higgs boson decays.
Therefore, any squark loop which causes 
an increase (decrease) in $\Gamma(h \to \gamma\gamma)$
will cause a proportionally larger 
decrease (increase) in $\Gamma(h \to gg)$.


\subsubsection{MSSM Higgs total decay width}

The total Higgs decay width in the MSSM is given, similarly to the SM,
by the sum of all the Higgs partial decay widths, {\it i.e.},
$\Gamma_{\textnormal{tot}} = \Gamma_{b\bar{b}} + \Gamma_{WW} + \Gamma_{ZZ} +
\Gamma_{\tau\bar{\tau}}$. Other partial decay widths into SM particles are much
smaller and can safely be neglected. As per decays into SUSY states,
we assume that the lightest
neutralino is heavy enough, so we do not have invisible
decay channels with large rates.  In the SM with a 125 GeV Higgs
mass, these partial decay widths are given by $\Gamma_{b\bar{b}} =
2.4\times 10^{-3}$ GeV, $\Gamma_{WW}= 8.8 \times 10^{-4}$ GeV,
$\Gamma_{ZZ}= 1.0 \times 10^{-4}$ GeV and
$\Gamma_{\tau\bar{\tau}}=2.4 \times 10^{-4}$ GeV.

In the MSSM, when $m_h\approx125$ GeV, this width is dominated by the
partial width to $b\bar b$, $\Gamma(h\to b\bar b)$, which is controlled 
by the bottom quark Yukawa coupling, $Y_b$.
In the SM, it is given by the expression $Y_b\equiv g_{hb\bar
b}=m_b/v$.

At large $\tan \beta$, sbottom-gluino and stop-chargino loops give
corrections to this Yukawa, which can be 
approximated by \cite{Carena:1999py}

\be%
Y_b \approx -\frac{m_b \sin \alpha}{ \cos \beta (1 + \Delta m_b) v}
\left( 1- \frac{\Delta m_b}{\tan \alpha \tan \beta}  \right)%
\ee%
where
\be%
\Delta m_b = \frac{2 \alpha_3}{3 \pi} m_g \mu \tan \beta I(m_{\tilde{b}_1}^2,m_{\tilde{b}_2}^2, \vert m_{\tilde{g}} \vert^2)
+ \frac{\vert h_t \vert^2}{16 \pi ^2} A_t \mu \tan \beta I(m_{\tilde{t}_1}^2,m_{\tilde{t}_2}^2, \vert \mu \vert^2)
\label{eq:dmb}
\ee%
and
\be%
h_t = \frac{m_t}{v\sin \beta} %
\ee%
with $\alpha_3$,  $m_{\tilde g}$ and $A_t$ being the SUSY-QCD
constant, gluino mass and top quark trilinear parameter,
respectively,
and where the loop function $I(a,b,c)$ is defined as %
\be%
I(a,b,c) = \frac{a b \ln(a/b) + b c \ln(b/c) + c
a \ln(c/a)}{(a-b)(b-c)(a-c)}.%
\ee%

$I(a,b,c)$ is a positive definite function, therefore with positive $m_g$, 
$\mu$ and $A_t$, the correction $\Delta m_b$ is positive, and
$Y_b$ is reduced. In particular, we see that this correction is large for
large values of $\mu$.

As the total width of the Higgs is dominated by the partial width to $b \bar b$,
a reduction in $Y_b$ will lead to a reduction in both $\Gamma(h\to b\bar 
b)$ and $\Gamma_{\textnormal{tot}}$, with a subsequent universal increase in all other 
BRs and 
$\mu_{X,Y} = \kappa_X \times \kappa_Y \times \frac{\Gamma^{\rm SM}_{\rm tot}}{\Gamma^{\rm MSSM}_{\rm tot}}$
irrespectively of the production channel.

However, in the decoupling limit $M_A >> M_Z$, $\tan \alpha \rightarrow -\cot\beta$, therefore
\be%
\left( 1 - \frac{\Delta m_b}{\tan\alpha \tan\beta} \right) \rightarrow 
\left( 1 + \Delta m_b \right)%
\ee%
which along with $\tfrac {\sin \alpha}{\cos \beta} \rightarrow -1$
means that $Y_b$ reduces to its SM value.

Therefore, to have the possibility for some reduction of $Y_b$,
we also consider the parameter space with values of $M_A$ not too 
large (the quasi-decoupling regime), such that
$Y_b$ can be reduced to be below its SM value.
At the same time  the $M_A$ values should be large enough 
such that $M_h \approx 125$ GeV is possible.

\begin{figure}[htb]
\begin{center}
\epsfig{file=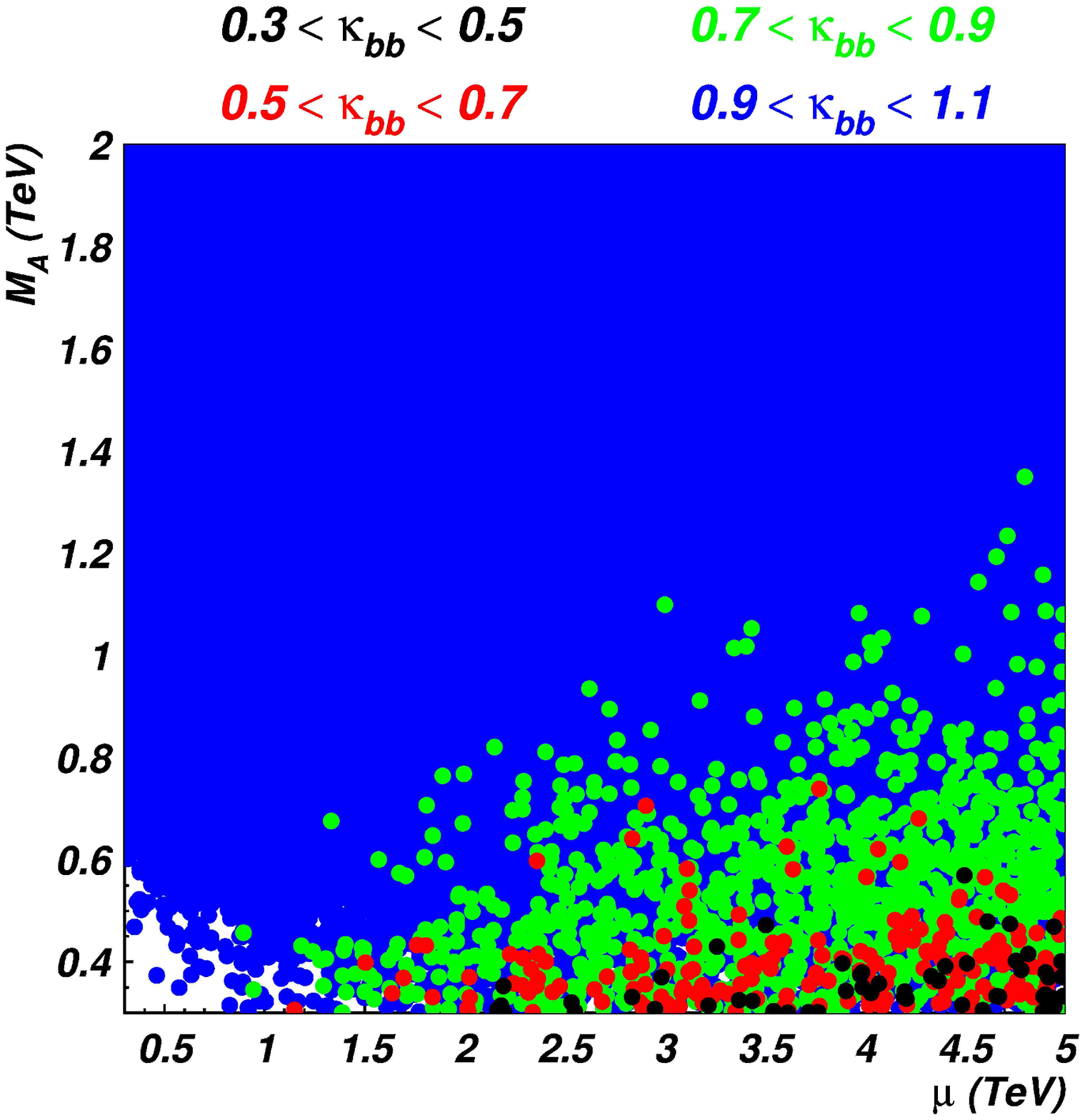,width=0.5\textwidth,angle=0}%
\epsfig{file=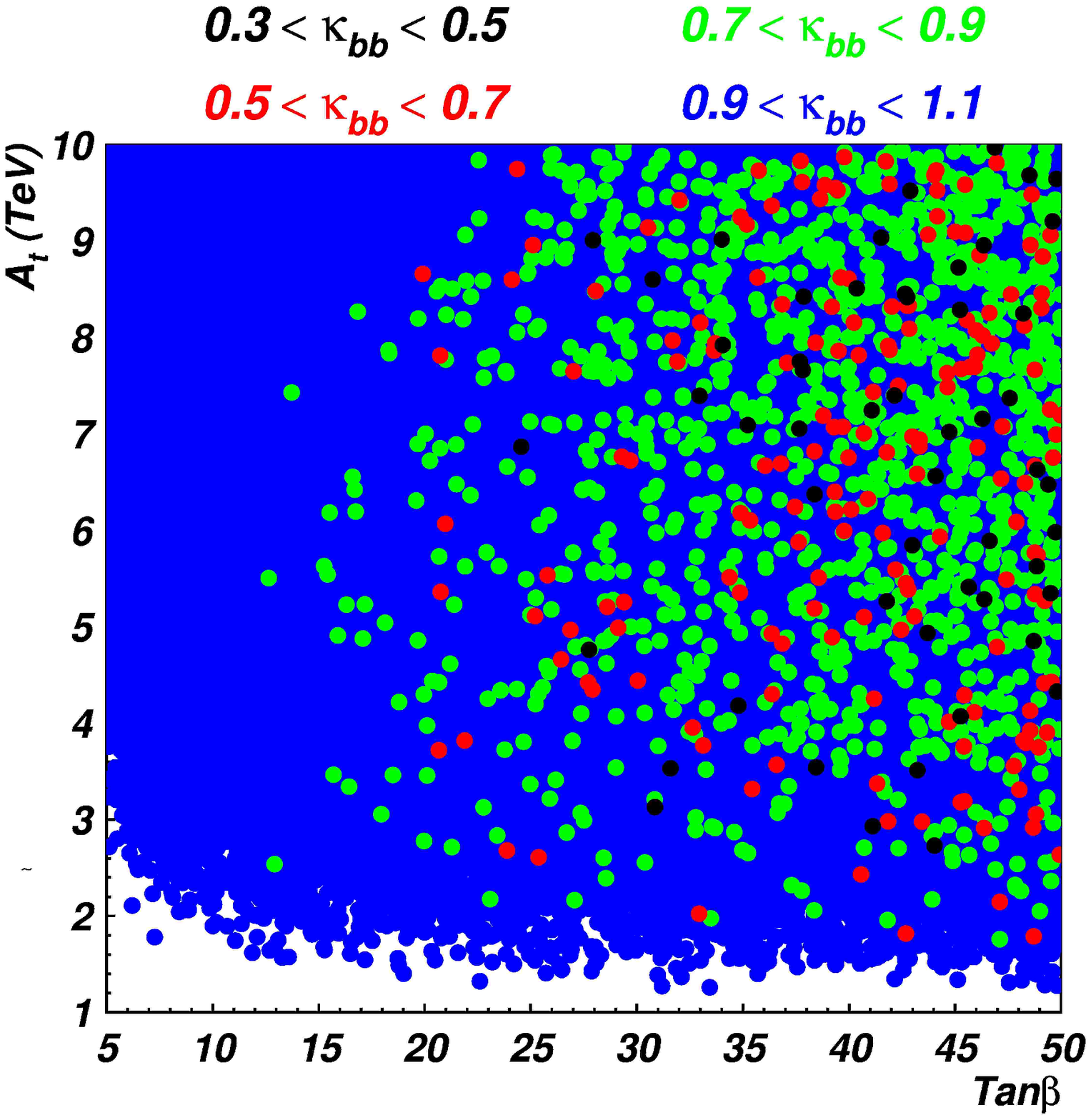,width=0.5\textwidth,angle=0}%
\caption{Results of the scan for $\kappa_{b \bar{b}}$
in the ($M_A, \mu$) (left)
and  ($A_t,\tan \beta$) (right) planes, respectively, where we have required 124 GeV $\le m_h \le$ 126 GeV.
} %
\label{yb_ma_mu}
\end{center}
\end{figure}

The results of our scan are presented in  Fig.~\ref{yb_ma_mu} where different values of $\kappa_{b \bar{b}}$ are  plotted in the ($M_A$,$\mu$) and  ($A_t,\tan \beta$) plane. 
From Fig.~\ref{yb_ma_mu}
one can see that large radiative SUSY corrections affecting $Y_b$
indeed are correlated to small values of $M_A$ and large values of $\mu$ (left frame)
as well as with large values of $\tan\beta$  (right frame).
These results are very consistent with eq. (\ref{eq:dmb})
which tells us that indeed large values of $\Delta m_b$ can be
achieved with large values of $\mu$ and/or $\tan \beta$.
At the same time, lower values of $M_A$ lead to the alteration of $Y_b$
at tree-level: one can see that even for $M_A\simeq 500$~GeV, as 
$\kappa_{bb}<0.5$ can be reached for sufficiently large values of $\tan\beta$ and $\mu$.
It is also worth mentioning that no obvious 
correlation of $\Delta m_b$ with values of $A_t$ can be seen in Fig.~\ref{yb_ma_mu}
since we require $M_h=125\pm 1$~GeV, and this drives in turn the value of $A_t$ to be 
around $\sqrt{6} M_{\rm SUSY}$, i.e. near the maximal mixing scenario. 
Thus 
$A_t I(m_{\tilde{t}_1}^2,m_{\tilde{t}_2}^2, \vert \mu \vert^2) \approx \sqrt{6}M_{\rm SUSY} I(m_{\tilde{t}_1}^2,m_{\tilde{t}_2}^2, \vert \mu \vert^2)$ 
which can be shown to decrease for large $M_{\rm SUSY}$, limiting its maximum contribution to $\Delta m_b$.

%% file: 03.2_stop.tex
\subsection{Stop quark effects}
\label{sec:stops}

As previously discussed, since $\tfrac{F_1}{F_0} \approx -20$, in order 
for stops to have a significant effect on 
($h \rightarrow \gamma\gamma$), $\hat{g}_{h\tilde{t}_1\tilde{t}_1}$ is required to be 
very large. Furthermore, a positive $\hat{g}_{h\tilde{t}_1\tilde{t}_1}$ coupling will decrease $\kappa_{\gamma\gamma}$ 
whilst a negative coupling will increase $\kappa_{\gamma\gamma}$.

In the decoupling limit, the Higgs coupling to the lightest stop
is given by \cite{Djouadi:2005gj}%
\bea%
\hat{g}_{h\tilde{t}_1\tilde{t}_1} =
 \frac{1}{2} \cos 2\beta \Big[\cos^2
\theta_{\tilde{t}} - \frac{4}{3} \sin^2 \theta_W \cos 2
\theta_{\tilde{t}} \Big]+ \frac{m_t^2}{M_Z^2} + \frac{1}{2} \sin
2 \theta_{\tilde{t}} \frac{m_t X_t}{M_Z^2},
\eea %
where $\theta_{\tilde{t}}$ is the stop mixing angle defined by
\begin{equation}
\sin 2 \theta_t = \frac{2m_t X_t}{m^2_{\tilde{t_1}} - m^2_{\tilde{t_2}}}
\end{equation}
and $X_t$ is given in terms of the the Higgs-stop trilinear
coupling as $X_t = A_t - \mu \cot\beta$.

The first term in the
equation is small compared to $\tfrac{m_t^2}{M_Z^2}$ and so can be
largely ignored. When $X_t$  is also small,
then  $\hat{g}_{h\tilde{t}_1\tilde{t}_1} \simeq \tfrac{m_t^2}{M_Z^2} > 0$ 
will lead to a  decrease of $k_{\gamma\gamma}$. For large
$X_t$, if $m_{\tilde{t}_1} < m_{\tilde{t}_2}$, it can be shown that $\sin 2 \theta_{\tilde{t}} \simeq -1$ and
the Higgs coupling to the lightest stop is strongly enhanced and negative. 
However, since  $m_h \approx 125$ GeV, the scenario  with light
stops requires that the Higgs mixing should be near maximal,
{\it i.e.}, $ X_t \approx \sqrt{6} M_{\rm SUSY}$, where
$M_{\rm SUSY}=\frac{1}{2}(m_{\tilde{t}_1} + m_{\tilde{t}_2})$.
Hence, we are not free to consider very large values of $X_t$
as an independent parameter.
In this case, one has%
\begin{equation}
\hat{g}_{h\tilde{t}_1\tilde{t}_1} \sim \frac{m_t^2}{M_Z^2} + \frac{3}{2}\frac{m_t^2}{M_Z^2} \frac{(m_{\tilde{t}_1} + m_{\tilde{t}_2})^2}{(m^2_{\tilde{t}_1} - m^2_{\tilde{t}_2})}.
\end{equation}
Thus, if $m^2_{\tilde{t}_2} \approx m^2_{\tilde{t}_1}$, it is possible to get
a very large Higgs coupling to stops. However, with a light stop, such that
$m^2_{\tilde{t}_1} \ll m^2_{\tilde{t}_2}$, one finds
\begin{equation}
\hat{g}_{h\tilde{t}_1\tilde{t}_1} \rightarrow \frac{m_t^2}{M_Z^2} - \frac{3}{2}\frac{m_t^2}{M_Z^2} = -\frac{1}{2}\frac{m_t^2}{M_Z^2}.
\end{equation}
Therefore  $\hat{g}_{h\tilde{t}_1\tilde{t}_1}$ is both negative (making the overall stop loop
contribution of the same sign as the $W$ loop), thereby increasing
$k_{\gamma\gamma}$, and fixed, which limits the overall 
contribution to ($h \rightarrow \gamma\gamma$) of a stop loop of a particular mass.
Small deviations from this prediction should be expected as in practice we are only
requiring \textit{near} maximal mixing.

\begin{figure}[htbp]
\epsfig{file=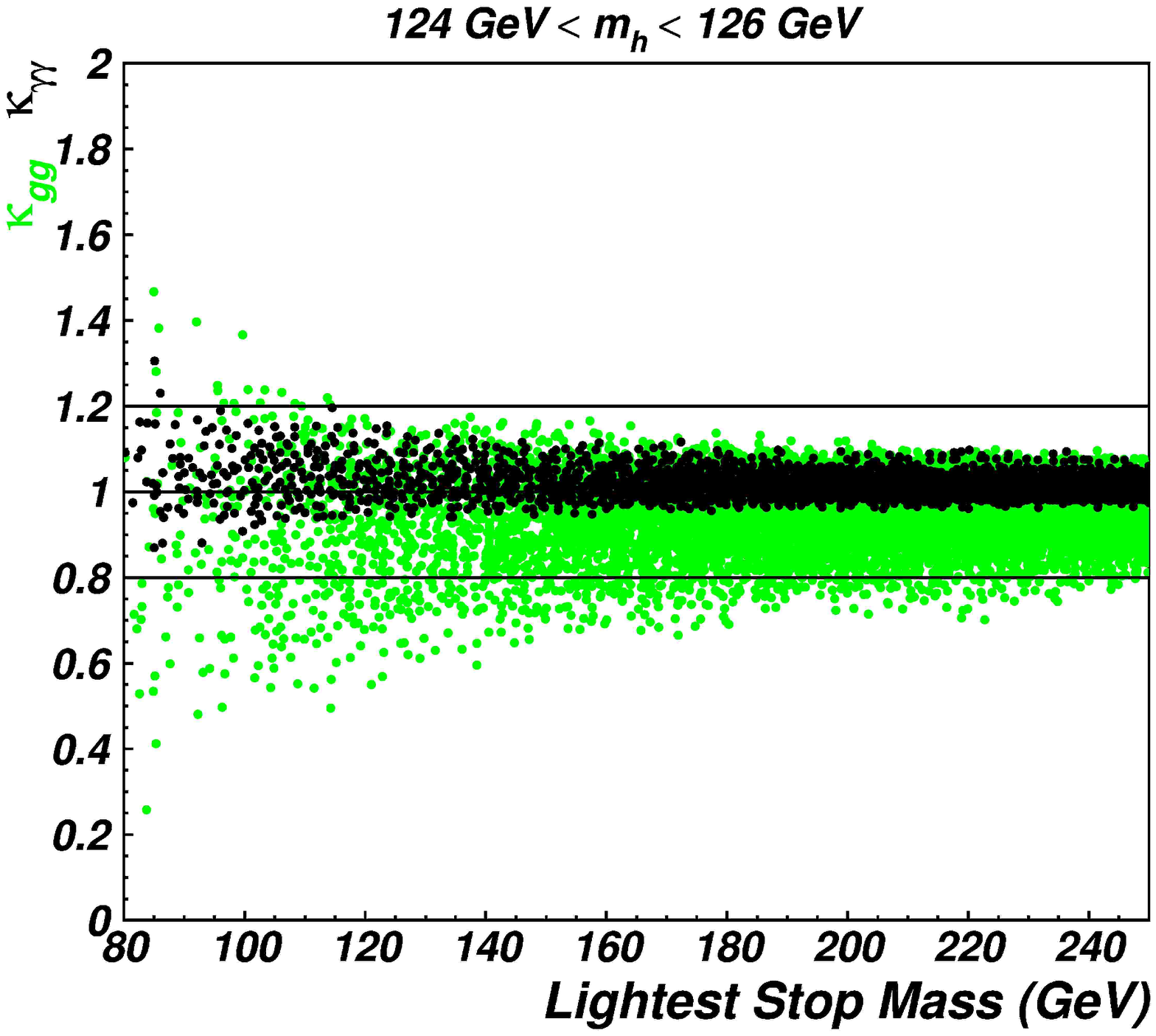,width=0.475\textwidth,angle=0}%
\epsfig{file=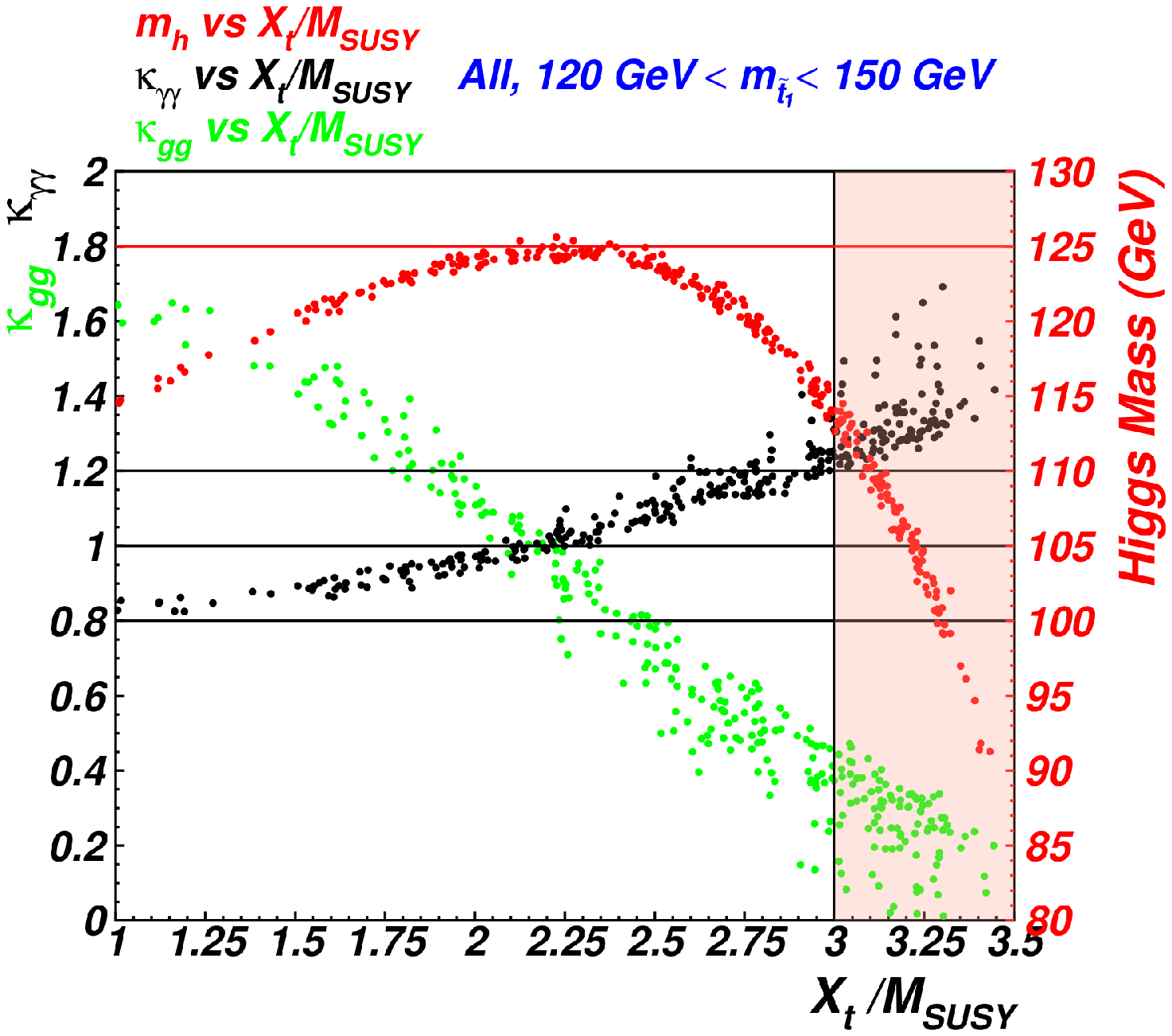,width=0.525\textwidth,angle=0}%
\\
\hspace*{0.2\textwidth}(a)\hspace*{0.5\textwidth}(b)
\caption{
Left: $\kappa_{\gamma\gamma}$ (black) and $\kappa_{gg}$ (green) as a function of lightest stop mass
for 124 GeV $< m_h <$ 126 GeV.
(b) $\kappa_{\gamma\gamma}$ (black), $\kappa_{gg}$ (green) and $m_h$ (red) as functions of
$\frac{X_t}{M_{\rm SUSY}}$ for $120$ GeV $\le m_{\tilde{t}_1} \le 150$ GeV. 
Cuts have been applied such that only points with a Higgs mass within 2 GeV 
of the maximum value for each value of $\frac{X_t}{M_{\rm SUSY}}$ are kept.
The pink-shaded window indicates the $X_t/M_{\rm SUSY} > 3$ 
region, where the majority of points do not pass the colour breaking minima conditions.
To isolate the influence of light stops, the following cuts are also applied to both plots:
$m_{H^{\pm}}$, $m_{\chi^{\pm}_{1,2}}$, $m_{\tilde{\tau}_{1,2}}$, $m_{\tilde{b}_{1,2}}$, $m_{\tilde{t}_2} > 300$ GeV.
}
\label{stop_1st}
\end{figure}

This is illustrated 
in Fig.~\ref{stop_1st}(a) where we present results for $\kappa_{\gamma
\gamma}$ as a function of the lightest stop mass, $m_{\tilde{t}_1}$, where 
124 GeV $< m_h <$ 126 GeV, for the scan described in Section II.
Together with  Fig.~\ref{stop_1st}(b) which presents $\kappa_{\gamma
\gamma}$ versus $X_t/M_{\rm SUSY}$ as well as $m_h$ versus $X_t/M_{\rm SUSY}$,
these two figures provide a clear illustration of the argument discussed above:
the effect of the stop is limited because of the correlation between 
$\hat{g}_{h\tilde{t}_1\tilde{t}_1}$ and the Higgs mass.
As can be seen from Fig.~\ref{stop_1st}(a), even $m_{\tilde{t}_1} \sim 120$ GeV
would lead to a modest increase of 
$\kappa_{\gamma\gamma} \approx 1.2$. At the same time, in Fig.~\ref{stop_1st} (b) 
which presents $\kappa_{\gamma\gamma}$, $\kappa_{gg}$ and $m_h$
as functions of $\frac{X_t}{M_{\rm SUSY}}$,
we can see that the effect of light stops on 
$\kappa_{\gamma\gamma}$ and $\kappa_{gg}$
could be much larger if $m_h$ was not limited to be in the 124--126 GeV mass window:
outside of this window $\kappa_{\gamma\gamma}$ could be as large as 1.8 
for a stop quark mass of about 120 GeV, however the majority of points with
$X_t/M_{\rm SUSY} > 3$ do not pass the colour breaking minima conditions discussed in
Section II, limiting the maximum $\kappa_{\gamma\gamma}$ to around 1.5.

In this figure it is clear that, in order to get $m_h \approx 125$ GeV, $X_t/M_{\rm SUSY}$ has to be
near its maximal mixing value of $\sqrt{6} \approx 2.4$. In this region 
there is little variation in $k_{\gamma \gamma}$ and we are limited
to $0.9 \lesssim k_{\gamma\gamma} \lesssim 1.2$.

\begin{figure}[htb]
\epsfig{file=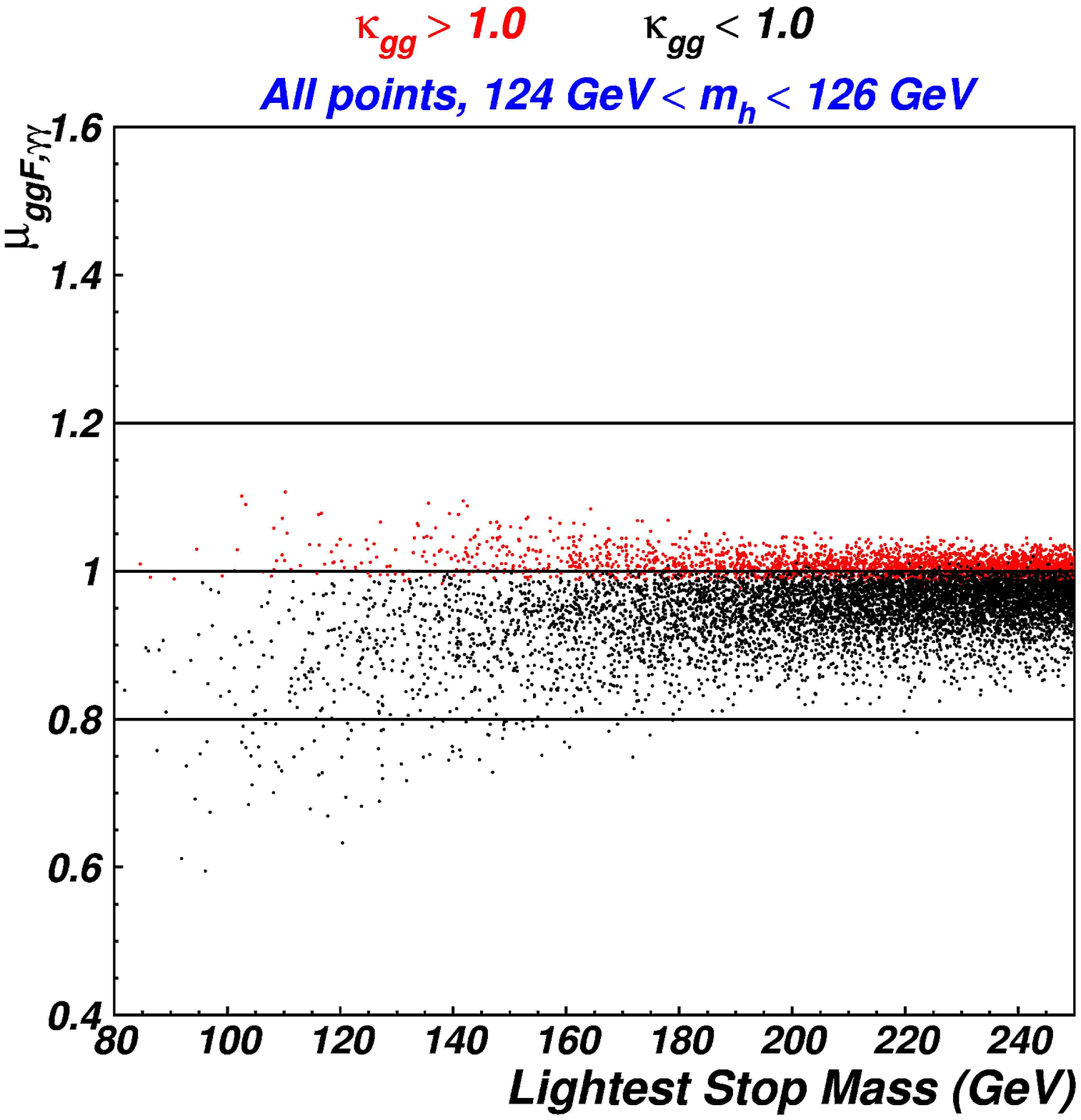,width=0.325\textwidth,angle=0}%
\epsfig{file=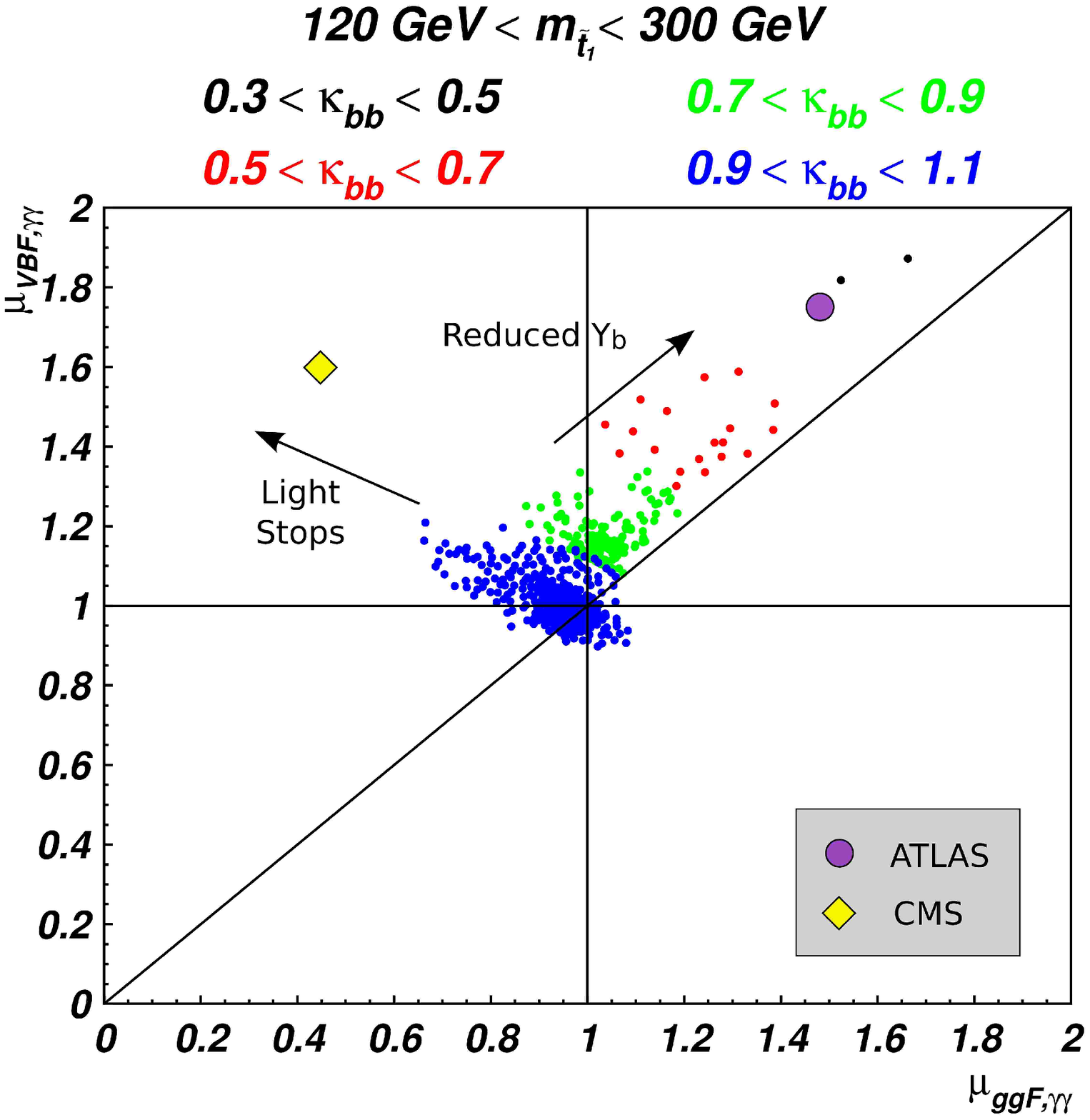,width=0.35\textwidth,angle=0}%
\epsfig{file=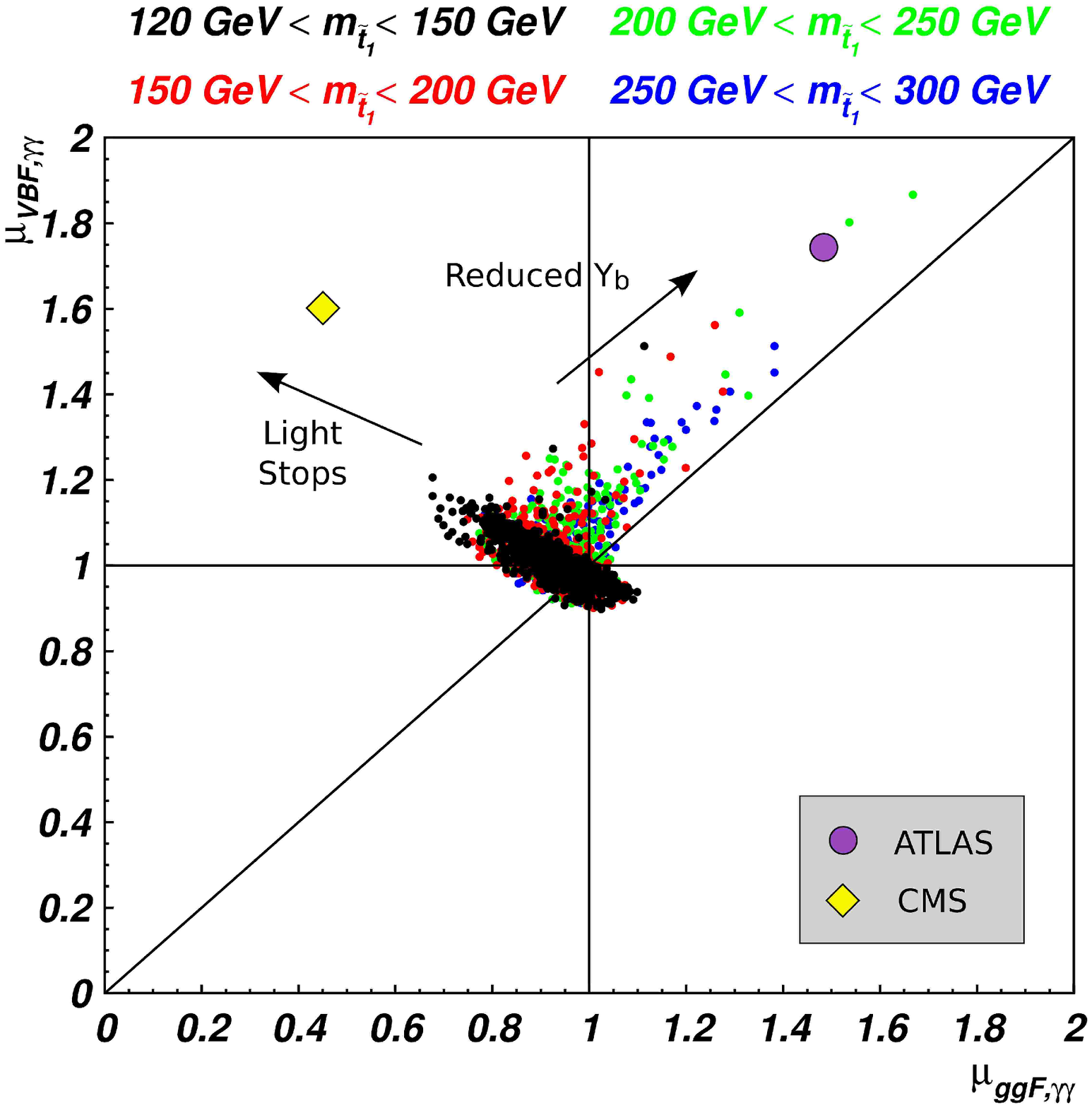,width=0.325\textwidth,angle=0}%
\\
\hspace*{0.1\textwidth}(a)\hspace*{0.25\textwidth}(b)(i)\hspace*{0.25\textwidth}(b)(ii)
\caption{
(a) $\mu_{ggF,\gamma \gamma}$ vs lightest stop mass for $\kappa_{gg} > 1$ (red) and $\kappa_{gg} \le 1$ (black).
We have cut for $0.98 \le \kappa_{bb} \le 1.02 $ to remove the possible effect of a reduced
$\Gamma_{hb\bar{b}}$.
(b) Each point of the scan with $120$ GeV $\le m_{\tilde{t}_1} \le 300$ GeV
is plotted on the ($\mu_{VBF},\mu_{ggF}$) plane, with colours to indicate
(i) different values for $\kappa_{bb}$,
(ii) different $m_{\tilde{t}_1}$ masses.
The results from ATLAS (purple circle) and CMS (yellow diamond) are indicated for comparison.
In all the plots, $124$ GeV $\le m_h \le 126$ GeV,
and to isolate the influence of light stops the following cuts are also applied:
$m_{H^{\pm}}$, $m_{\chi^{\pm}_{1,2}}$, $m_{\tilde{\tau}_{1,2}}$, $m_{\tilde{b}_{1,2}}$, $m_{\tilde{t}_2} > 300$ GeV.
} %
\label{stop_2nd}
\end{figure}

Let us consider the overall effect of light stops on 
$\mu_{ggF,\gamma\gamma} = k_{gg} \times k_{\gamma \gamma} \times k_h^{-1}$ via its effects
on $k_{\gamma \gamma}$ and $k_{gg}$ in the parameter space  where the total width is
close to the SM one. From
Fig.~\ref{stop_1st} (b) we would expect that, in general, either
$k_{\gamma \gamma}$ is increased with a relatively larger decrease in $k_{gg}$,
causing an overall decrease in $\mu_{ggF,\gamma\gamma}$, or $k_{\gamma \gamma}$ 
is decreased with a relatively larger increase in $k_{gg}$, causing 
an overall increase in $\mu_{ggF,\gamma\gamma}$. This is demonstrated in 
Fig.~\ref{stop_2nd}(a), where we see that 
(other than for a few points very
near $\mu_{ggF,\gamma\gamma}=1$ where other factors such as small changes in 
the total width play a role)  we have $\mu_{ggF,\gamma\gamma} > 1$ when 
$k_{gg} > 1$ (red) and vice versa (black). This means that if the total width of the Higgs boson 
is unchanged, then stop loops alone can produce a universal increase
in all decay channels ($\mu_{ggF,Y} > 1$) via increasing the $ggF$ 
production channel but will not produce an isolated increase in 
($h \rightarrow \gamma\gamma$), as this will always be cancelled by a
relatively larger decrease in $ggF$ production.

We are naturally lead to consider the possibility of counteracting
the effect of a reduced $\kappa_{gg}$ caused by light stops by reducing
$\Gamma_{b\bar{b}}$ as discussed in Section IIIA.3.
This would mean that when the stop coupling is negative, producing an 
increase in $k_{\gamma \gamma}$ and bigger relative decrease of  $\kappa_{gg}$, 
the BRs in all channels other than $b \bar b$ can be increased such that the
overall value for $\mu_{ggF,Y}$ remains $\approx 1$. In this scenario,
$\mu_{VBF,\gamma\gamma} > \mu_{ggF,\gamma\gamma}$ as the $VBF$ channel will 
be increased by both $k_{\gamma \gamma} > 1$ and the increased BR to photons 
from the reduced total width, without the reduced production
rate of the $ggF$ channel. This is demonstrated in Fig.~\ref{stop_2nd} (b) 
where the effects of light stops and reduced $\Gamma_{b\bar{b}}$ (via a
reduction in the bottom Yukawa coupling) are combined together and the
resulting $\mu_{VBF,\gamma\gamma}$ and  $\mu_{ggF,\gamma\gamma}$ 
values along with current best fit CMS and ATLAS data are plotted. 
One can see  in Fig.~\ref{stop_2nd} (b)(i) that the smaller the $\kappa_{bb}$ values,
the larger the {\it universal} $\mu_{VBF,\gamma\gamma}$ and $\mu_{ggF,\gamma\gamma}$ 
alterations it will cause, while Fig.~\ref{stop_2nd} (b)(ii) clearly demonstrates how decreased 
stop quark masses lead to an increase of the {\it non-universal} 
alteration of these couplings, which can be expressed through the $\mu_{VBF,\gamma\gamma}/\mu_{ggF,\gamma\gamma}$ ratio. 

In this scenario, the same situation takes place for
all other decay channels with the exception of the decay to bottoms, 
($\mu_{VBF,WW/ZZ/\tau\tau/\gamma\gamma} > \mu_{ggF,WW/ZZ/\tau\tau/\gamma\gamma}$), as will be discussed 
further in Section III.E.
Furthermore, in this region, the di-photon decay channel can
be increased by a factor of up to 1.2 relative to the other decay channels.
\begin{figure}[htb]
\begin{center}
\epsfig{file=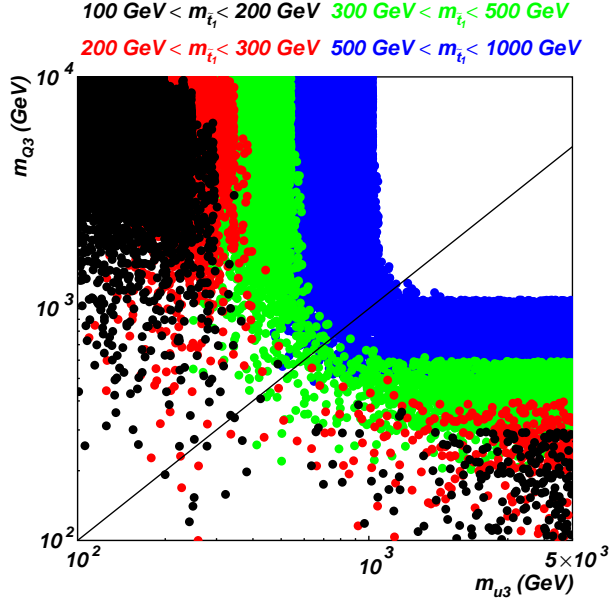,width=0.5\textwidth,angle=0}%
\\
\caption{
Different values for $m_{\tilde{t}_1}$ in the $M_{U3}$ versus $M_{Q3}$ plane. 
We have required $124$ GeV $\le m_h \le 126$ GeV.
} %
\label{MU3vsMQ3}
\end{center}
\end{figure}

Of note, the area of parameter space with light stops and a Higgs mass of $\approx125$ GeV
is relatively small, because a heavy $M_{\rm SUSY}$ is preferred to give large logarithmic
parts to the radiative corrections to the Higgs Mass. 
The scenario where $M_{U3} \sim M_{Q3} \gtrsim 300$ GeV requires fine-tuning 
of the stop mixing parameter $X_t$ in order to achieve a lightest stop  with mass $\le 300$ GeV.
However, as $X_t$ is fixed by the near maximal mixing requirement ($X_t \approx \sqrt{6} M_{\rm SUSY}$),
this is not possible. Hence, the area of parameter space with
a $\approx125$ GeV lightest Higgs mass and light stops (with mass $\le 300$ GeV) is where
$M_{U3} \ll M_{Q3}$, generally with $M_{U3} \le 300$ GeV and
$M_{Q3} \gtrsim 2$ TeV. (In this region it is easy to show that $m_{\tilde{t_1}} \approx M_{U3}$ 
and $m_{\tilde{t_2}} \approx M_{Q3}$). This explains the reason for choosing the reduced
range of $M_{U3}$ described in Section II for the additional scan.
The relationship between $M_{U3}$, $M_{Q3}$ and the lightest stop mass is shown
in Fig.~\ref{MU3vsMQ3}, exemplifying a strong correlation between $m_{\tilde{t_1}}$ and $M_{U3}$
as discussed.

\begin{table}[htb]
\begin{tabular}{|c|c|c|c|}
 \hline\hline
  ~~~~Parameter~~~~ 		& ~~~Benchmark 1~~~ 	& ~~~Benchmark 2~~~   & ~~~Benchmark 3~~~ \\
  \hline
  $\tan \beta$ 			& 37     		& 48    		& 44 \\
  $\mu$ 			& 300 GeV               & 2 TeV          	& 400 GeV \\
  $M_{H^\pm}$ 			& 1.7 TeV		& 1 TeV  		& 750 GeV  \\
  $M_{Q3}$       		& 2.5 TeV       	& 2.5 TeV		& 1.3 TeV \\  
  $M_{U3}$      		& 165 GeV  	        & 230 GeV		& 320 GeV \\  
  $M_{D3}$      		& 11 TeV          	& 12 TeV		& 7 TeV \\  
  $M_{L3}$      		& 4 TeV       	& 3 TeV		        & 2 TeV \\  
  $M_{E3}$      		& 1.2 TeV       	& 500 GeV		& 5 TeV \\  
  $M_3$     		        & 1.9 TeV       	& 3.2 TeV		& 2 TeV \\  
  $M_2$     		        & 3 TeV          	& 3 TeV  		& 3 TeV \\  
  $M_1$     		        & 125 GeV          	& 172 GeV  		& 250 GeV \\  
  $A_t$ 		        & 3.1 TeV               & 3.6 TeV   		& 2.1 TeV \\
  $A_b$ 			& 5.5  TeV                & 100 GeV		& 7 TeV  \\
  $A_\tau$ 			& 500 GeV               & 0 GeV	        	& 2.5 TeV \\
\hline
  $m_{\tilde{t}_1}$    		& 125 GeV 	        & 177 GeV		& 254 GeV \\  
  $m_{\tilde{\chi}_1^0}$  		& 121 GeV 	        & 172 GeV		& 245 GeV \\  
  $m_h$           		& 124.1 GeV       	& 124.0 GeV		& 124.2 GeV \\  
\hline
  $\frac{\mu_{VBF,\gamma\gamma}}{\mu_{ggF,\gamma\gamma}}$  & $\frac{1.11}{0.78} = 1.42$ & $\frac{1.65}{1.16} = 1.42$ & $\frac{1.08}{0.80} = 1.35$ \\
\hline
  $\kappa_{gg}$                  & $0.71$                   & $0.70$                   & $0.74$ \\
  $\kappa_{\gamma\gamma}$          & $1.10$                   & $1.10$                   & $1.08$ \\
  $\kappa_{bb}$                  & $1.01$                   & $0.55$                   & $1.04$ \\
  $\kappa_h$                    & $0.99$                   & $0.67$                   & $1.01$ \\
  $\mu_{ggF,bb}$                  & $0.72$                   & $0.58$                   & $0.76$ \\
  $\Delta^{-1}$                  & $4.6 \%$                 & $0.1 \%$                 & $2.6 \%$ \\
\hline\hline
\end{tabular}
\caption{Benchmark points with light stops and $\tfrac{\mu_{VBF}}{\mu_{ggF}} > 1$}\label{benchmarks}
\end{table}

In Tab. \ref{benchmarks}, we give three different benchmark points for scenarios where
light stops give rise to $\tfrac{\mu_{VBF}}{\mu_{ggF}} > 1$, where we have also included a value 
for the minimum fine-tuning for each of the benchmark points.

Our fine-tuning parameter
is based on the electroweak fine-tuning parameter (see  e.g.  \cite{Baer:2012cf} for details). 
This value is derived
by noting that the minimisation condition for the Higgs potential gives rise to the 
equation for the Z-boson mass,
\begin{equation}
\frac{M_Z^2}{2} = \frac{m^2_{H_d} + \Sigma^d_d - \left( m^2_{H_u} + \Sigma^u_u \right)\tan^2 \beta}{\tan^2 \beta - 1} - \mu^2
\end{equation}
where $\Sigma^u_u$ and $\Sigma^d_d$ are the radiative corrections to $m^2_{H_u}$ and $m^2_{H_d}$. 
The electroweak fine-tuning parameter is then defined as
\begin{equation}
\Delta_{EW} \equiv max_i (C_i)/(M_Z^2/2)
\end{equation}
where $C_{H_u} = |-m^2_{H_u} \tan^2 \beta / (\tan^2 \beta - 1)|$, with analogous definitions
for $C_{H_d}$, $C_\mu$, $C_{\Sigma^u_u}$  and $C_{\Sigma^d_d}$. If $\tan \beta$ is moderate or large, we have
\begin{equation}
\frac{M_Z^2}{2} \approx -\left( m^2_{H_u} + \Sigma^u_u \right) - \mu^2.
\end{equation}\\
Taking into account that $\Sigma^u_u$ is not defined  if our starting point is the theory at the EW scale 
(rather than the GUT scale) the measure of fine-tuning 
\begin{equation}
\Delta \equiv |\mu^2|/(M_Z^2/2)
\end{equation}
gives a {\it minimum} value for $\Delta_{EW}$, which could be larger if there is a 
large cancellation between $m^2_{H_u}$ and $\Sigma^u_u$ as discussed by Baer et. al. in \cite{Baer:2012cf}.
Keeping this in mind we will be using this definition of fine-tuning in our paper.

We see that for Benchmark 
points 1 and 3, $\Delta^{-1} \sim 2-5 \%$, as there is no requirement for a large $\mu$.
However Benchmark 2 has a reduced $k_{bb}$ which requires a large $\Delta m_b$ and hence 
a large $\mu$, leading to a larger minimum fine-tuning, with $\Delta^{-1} \sim 0.1 \%$.

We should
also mention here that the light stop scenario has been discussed recently in connection to EWBG~\cite{Curtin:2012aa,Carena:2012np}. The latter requires $A_t\lesssim M_{Q3}/2$ (or  $X_t/M_{\rm SUSY}\lesssim \frac{1}{\sqrt{2}}$) 
in order to achieve 
a strong phase transition  and therefore is not  realised  in the maximal mixing scenario which we consider in our paper.
Remarkably, for these values of  $X_t/M_{\rm SUSY}$, the 125 GeV Higgs mass can be only achieved for extremely large values of
$M_{Q3}\simeq 10^6$~TeV, as shown in~\cite{Carena:2012np}. At the same time, $X_t/M_{\rm SUSY}\lesssim \frac{1}{\sqrt{2}}$
leads to $\hat{g}_{h\tilde{t}_1\tilde{t}_1}>0$ and therefore to a constructive interference of the light
stops inside $ggF$ production. It was suggested in \cite{Carena:2012np} that the overall enhancement
for  $ggF$  production coming from the light stop can be compensated by a significant invisible Higgs boson decay into light 
neutralinos.

We would also like to note that the scenario with an altered  $Y_b$ coupling that we consider in this paper
suggests an alternative solution to the problem of how to compensate the enhancement of $ggF$ production due to light
stops. Analogously to the parameter space region where  $Y_b$ is decreased, there is another
where  $Y_b$ is increased, which is realised in the $\mu<0$ region. In this case, the  enhancement of $ggF$ 
production is compensated by the  respective decrease of the $\gamma\gamma/ZZ/WW/\tau\tau$  decay rates
while  the BR($h\to b\bar{b}$) will be increased. Since, at the moment, the LHC is not quite sensitive
to  the $h\to b\bar{b}$ signature, this scenario is perfectly viable.


%% file: 03.3_sbottom.tex
\subsection{Sbottom quark effects}

Similarly to stops, light sbottoms loops may also alter Higgs production via $ggF$ 
and decay to di-photons. However, there are some important differences between the sbottom loop 
contribution and the stop loops.

In the decoupling limit, the Higgs coupling to sbottoms is given by%
\bea%
\hat{g}_{h\tilde{b}_1\tilde{b}_1} = \cos 2 \beta \left(-\frac{1}{2} \cos^2
\theta_{\tilde{b}} +  \frac{1}{3}\sin^2 \theta_W \cos 2
\theta_{\tilde{b}}\right)  + \frac{m_{b}^2}{M_Z^2}
+ \frac{m_{b}X_b}{2 M_Z^2} \sin 2 \theta_{\tilde{b}},%
\label{hbb_coupling}
\eea %

where $\theta_{\tilde{b}}$ is the sbottom mixing angle defined by
\begin{equation}
\sin 2 \theta_{\tilde{b}} = \frac{2m_b X_b}{m^2_{\tilde{b_1}} - m^2_{\tilde{b_2}}}
\end{equation}

and
$X_b = A_b - \mu \tan \beta$.

The first major difference with respect to the stop case is that this coupling does not have any dependence on $X_t$ and,
hence, is not constrained by the requirement of $m_h \approx 125$ GeV. In particular, when
$m_{\tilde{b}_1} < m_{\tilde{b}_2}$ with a large positive $\mu$ and $\tan \beta$, leading to a large
negative $X_b$, it can be shown than $\sin 2 \theta_{\tilde{b}} \sim 1$. 
The last term in eq.~(\ref{hbb_coupling}) therefore dominates, giving 
\begin{equation}
\hat{g}_{h\tilde{b}_1\tilde{b}_1} \simeq \frac{m_{b}X_b}{2 M_Z^2}
\end{equation}
and leading to a large negative coupling due to the negative $X_b$. 
As the Higgs-sbottom coupling ultimately depends on $X_{b}^2$, via the $\sin 2 \theta_{\tilde{b}}$ term, 
it is also possible to get
a large negative coupling if $\mu$ is large and negative such that $X_{b}$ is large and positive. 
However, as we are interested in the parameter space where $Y_b$ has the possibility of being small, 
which requires a positive $\mu$ (see Section III.A), we have only considered positive $\mu$.

The second important difference with respect to the stop case
is that $N_{c,b} Q_b^2 = 1/3$ versus $N_{c,t} Q_t^2 = 4/3$ for stops.
This will not affect gluon fusion, but means that the sbottom mass will need to be $\tfrac{1}{2}$
that of a stop mass with the same coupling strength to Higgs bosons to produce the same
alteration in decay to photons.

The result of the two factors discussed above is that, firstly, 
due to the opposing effects of the sbottom coupling having a larger maximum magnitude compared to
stops, but the sbottom loop effects of decay to di-photons being suppressed by a factor of
$\tfrac{Q_t^2}{Q_b^2}$ compared to stops, the relative effects of sbottom loops on $\kappa_{\gamma \gamma}$
compared to stop loops is difficult to predict. 
Secondly, the maximum effect of sbottom loops on 
gluon fusion can be larger than that of a stop with the same mass,
because the coupling isn't constrained by the Higgs mass and can become larger in magnitude
than the stop coupling, while the loop contribution isn't constrained by a charge factor 
as is the case for di-photon decay.
\begin{figure}[htb]
\epsfig{file=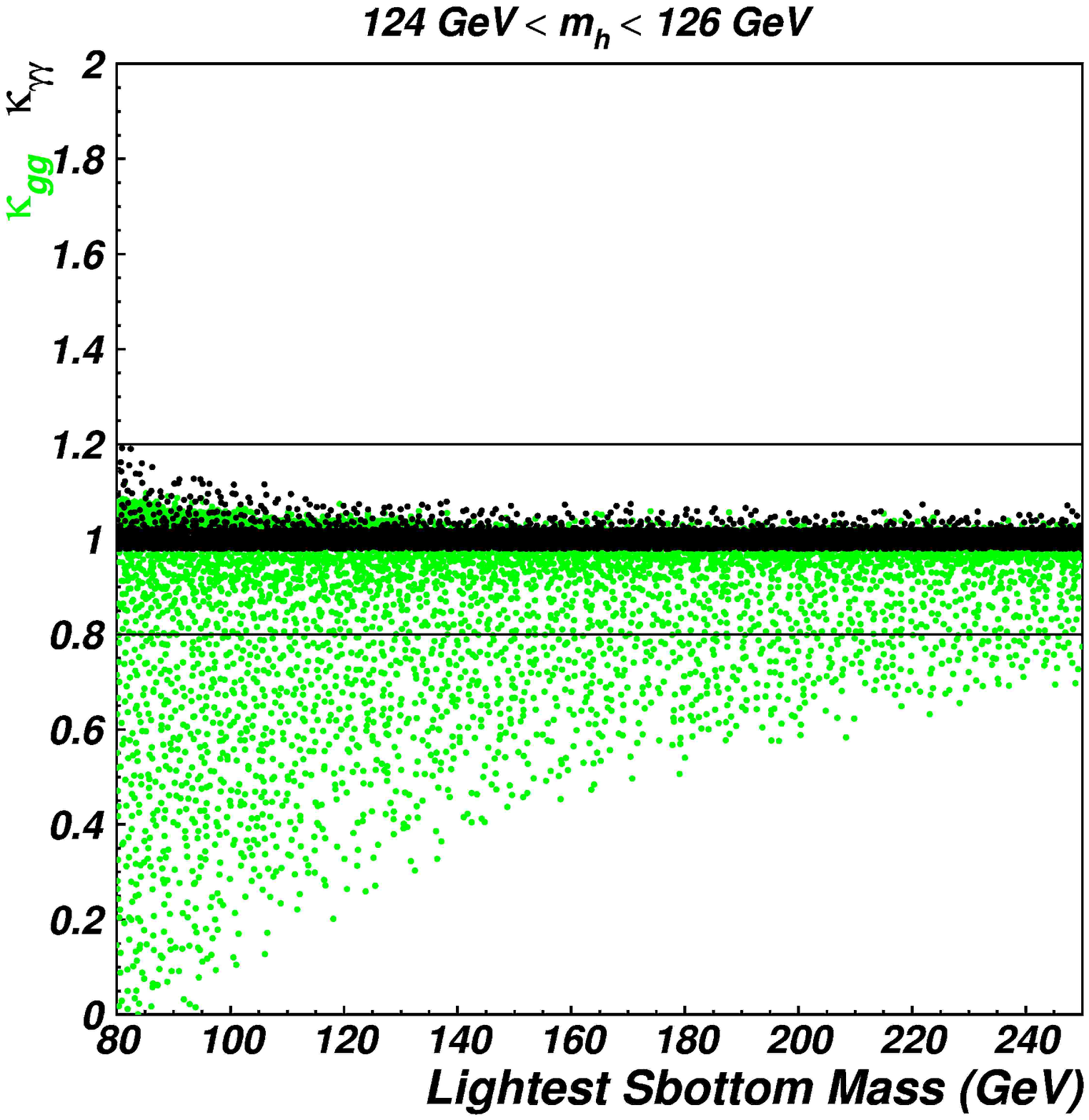,width=0.46\textwidth,angle=0}%
\epsfig{file=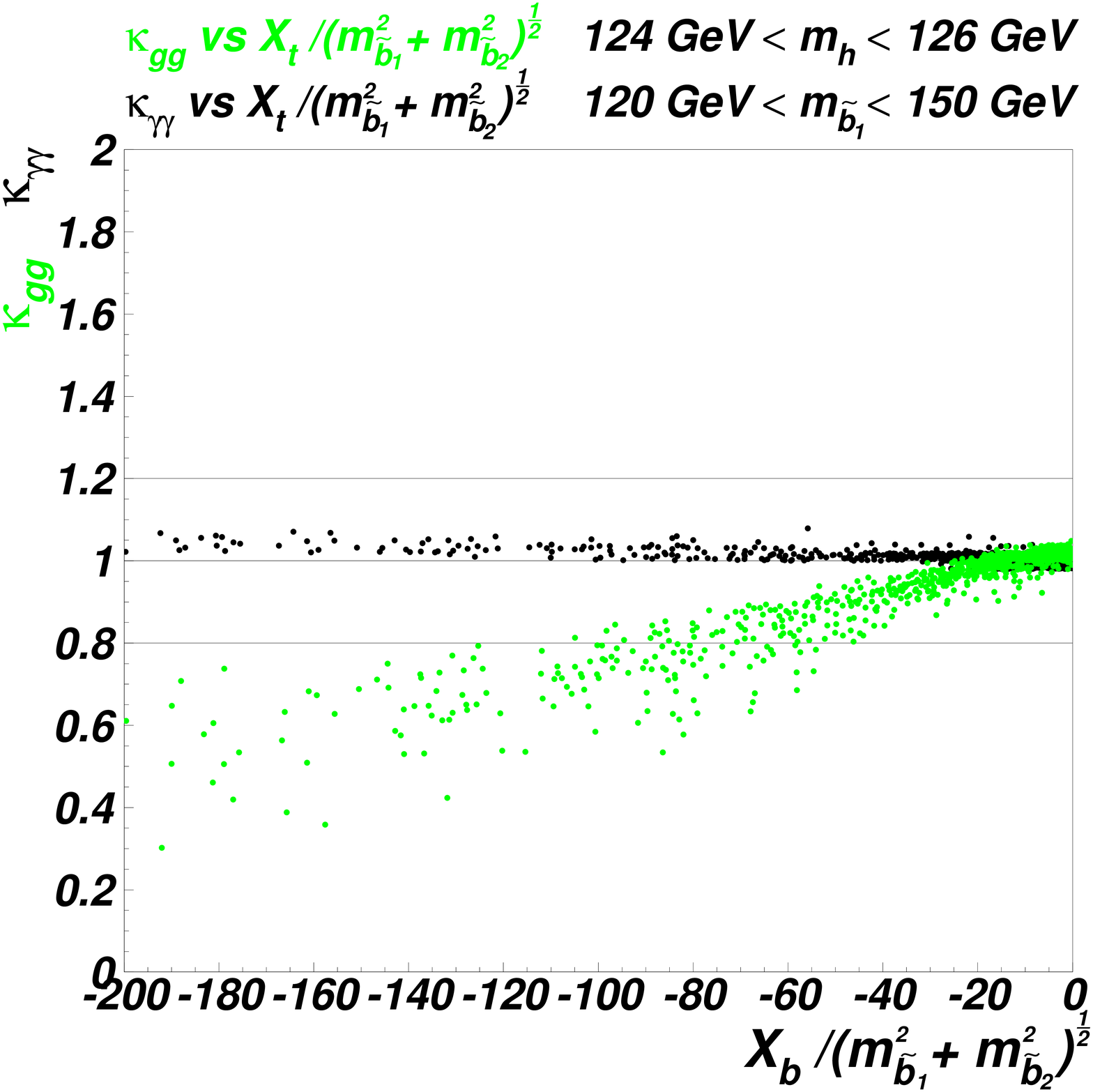,width=0.54\textwidth,angle=0}%
\\
\hspace*{0.2\textwidth}(a)\hspace*{0.5\textwidth}(b)
\caption{
(a) $\kappa_{\gamma\gamma}$ (black) and $\kappa_{gg}$ (green) as a function of lightest sbottom mass
for 124 GeV $< m_h <$ 126 GeV.
(b) $\kappa_{\gamma\gamma}$ (black) and $\kappa_{gg}$ (green) as functions of
$\frac{X_b}{(m_{\tilde{b}_1}^2 + m_{\tilde{b}_2}^2)}$ for $120$ GeV $\le m_{\tilde{b}_1} \le 150$ GeV.
To isolate the influence of light sbottoms, the following cuts are also applied:
$m_{H^{\pm}}$, $m_{\chi^{\pm}_{1,2}}$, $m_{\tilde{t}_{1,2}}$, $m_{\tilde{\tau}_{1,2}}$, $m_{\tilde{b}_2} > 300$ GeV.
} %
\label{sbottom_1st}
\end{figure}
\begin{figure}[htb]
\epsfig{file=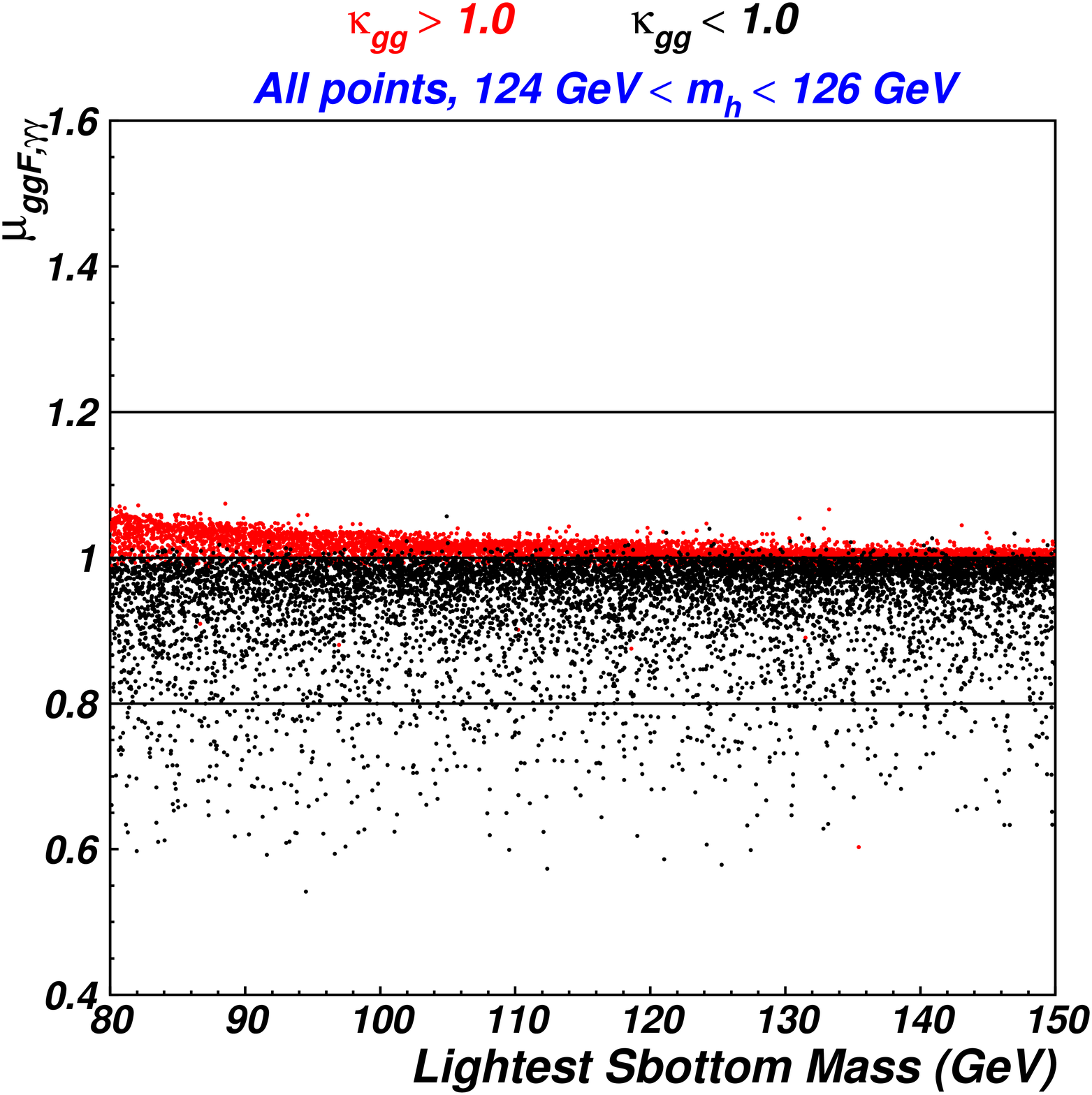,width=0.325\textwidth,angle=0}%
\epsfig{file=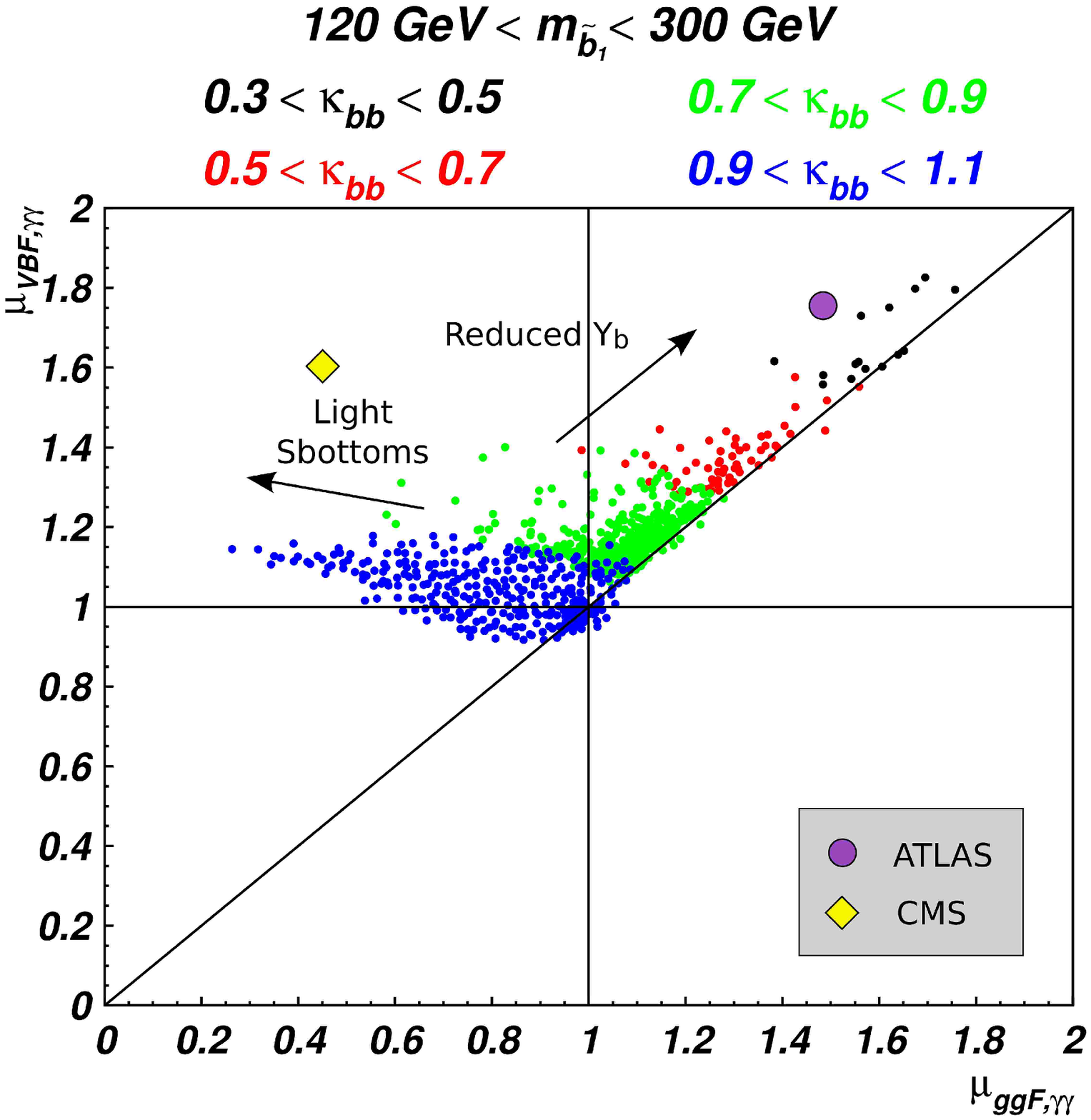,width=0.35\textwidth,angle=0}%
\epsfig{file=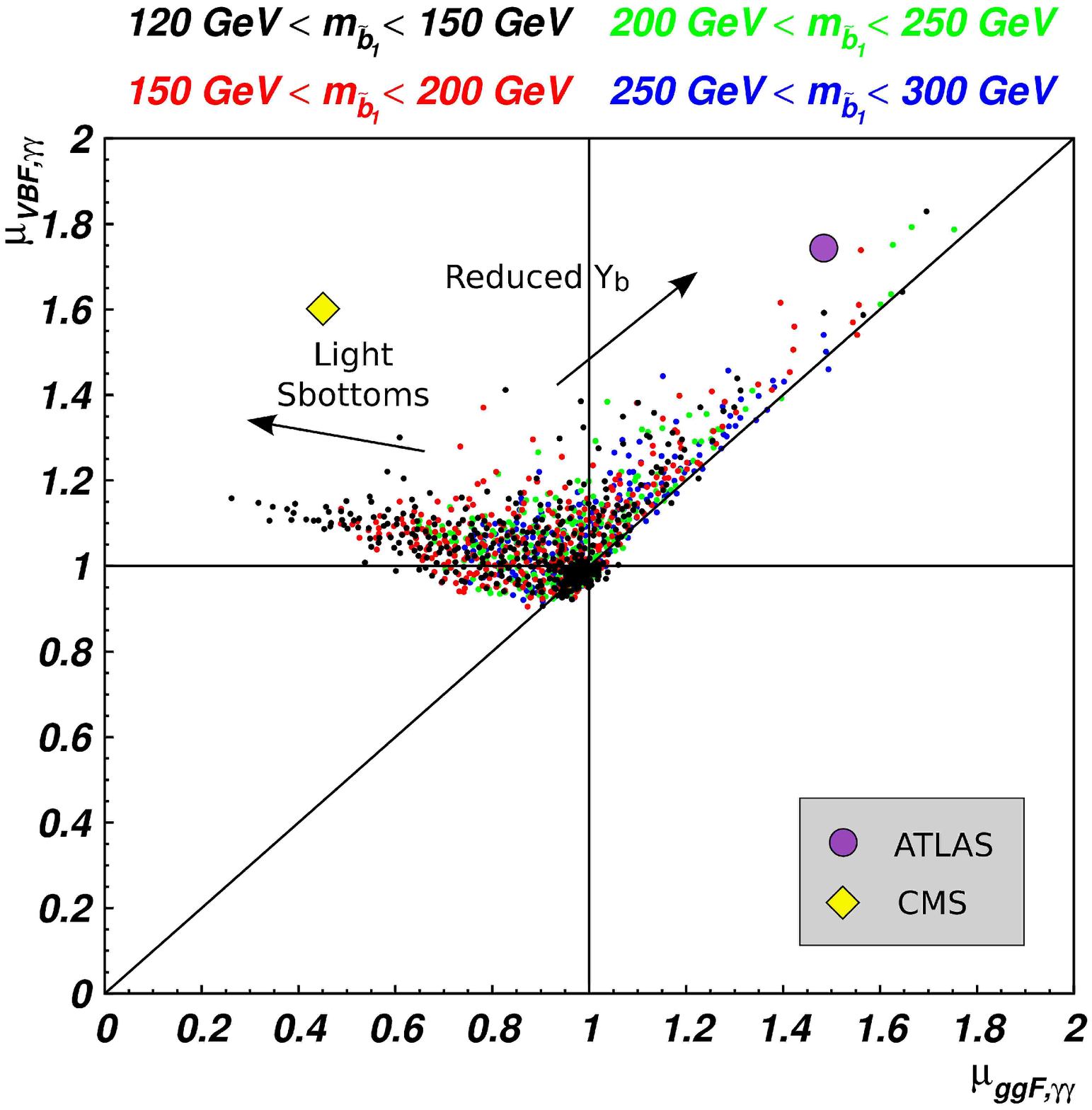,width=0.325\textwidth,angle=0}%
\\
\hspace*{0.1\textwidth}(a)\hspace*{0.25\textwidth}(b)\hspace*{0.25\textwidth}(c)
\caption{
(a) $\mu_{ggF,\gamma \gamma}$ vs lightest sbottom mass for $\kappa_{gg} > 1$ (red) and $\kappa_{gg} \le 1$ (black).
We have cut for $0.98 \le \kappa_{bb} \le 1.02 $ to remove the possible effect of a reduced
$\Gamma_{hb\bar{b}}$.
(b) Each point of the scan with $120$ GeV $\le m_{\tilde{b}_1} \le 300$ GeV
is plotted on the ($\mu_{VBF},\mu_{ggF}$) plane, with colours to indicate
(i) different values for $\kappa_{bb}$,
(ii) different $m_{\tilde{b}_1}$ masses.
The results from ATLAS (purple circle) and CMS (yellow diamond) are indicated for comparison.
In all the plots, $124$ GeV $\le m_h \le 126$ GeV,
and to isolate the influence of light sbottoms the following cuts are also applied:
$m_{H^{\pm}}$, $m_{\chi^{\pm}_{1,2}}$, $m_{\tilde{t}_{1,2}}$, $m_{\tilde{\tau}_{1,2}}$, $m_{\tilde{b}_2} > 300$ GeV.
} %
\label{sbottom_2nd}
\end{figure}

Both of these effects can be observed in Fig.~\ref{sbottom_1st}(a). We see
that the largest possible increase in $\kappa_{\gamma\gamma}$ (black) for a given sbottom mass
is smaller than that for a stop of the same mass, and only very light sbottoms
$\simeq 80$ GeV are able to produce $\kappa_{\gamma\gamma} \approx 1.2$ 
(compare to $m_{\tilde{t}_1} \approx 120$ GeV).
Also, as expected, $\kappa_{gg}$ (green) has a larger reduction for sbottoms compared to 
stops of similar mass, with $\kappa_{gg}$ as low as 0.7, for $m_{\tilde{b}_1} \approx 250$ GeV.

Fig.~\ref{sbottom_1st}(b) shows how $\kappa_{\gamma\gamma}$ (black) and $\kappa_{gg}$ (green)
depend on $\sqrt{\tfrac{X_b}{m_{\tilde{b}_1}^2 + m_{\tilde{b}_2}^2}}$ for 120 GeV $< m_{\tilde{b}_1} <$ 150 GeV,
confirming that the largest
deviations from the SM occur for large and negative $X_b$, and that the
effect on $\kappa_{gg}$ is much larger than for $\kappa_{\gamma\gamma}$.

The combined effect of the sbottom loops can be seen in Fig.~\ref{sbottom_2nd}(a), where
the lightest sbottom mass is plotted against $\mu_{ggF,\gamma\gamma}$. We see that 
for the majority of parameter space, $\mu_{ggF,\gamma\gamma}$ is suppressed, other than a
small region where it is increased due an increased $\kappa_{gg}$.
This small region of increased $\kappa_{gg}$ occurs for very small values of $X_b$ 
where the final term of eq.~(\ref{hbb_coupling}) does not dominate and the coupling
is small and positive.

If we consider the possibility of counteracting
the effect of a reduced $\kappa_{gg}$ by reducing
$\Gamma_{b\bar{b}}$ as we did for the stops, we find that we would expect sbottom
loops to have a similar effect to stop loops in the ($\mu_{VBF+VH},\mu_{ggF+ttH}$)
plane. As in the stop case, $VBF$ production channels will be unaffected by 
the gluon fusion rate, but the decays to all particles (other than sbottoms)
will still be increased by the reduction in $\Gamma_{b\bar{b}}$, such that
$\mu_{VBF,\gamma\gamma} > \mu_{ggF,\gamma\gamma}$.

This is demonstrated in Fig.~\ref{sbottom_2nd}(b)(i), which is analogous to 
Fig.~\ref{stop_2nd}(b)(i) for stops, where we have plotted  
$\mu_{VBF,\gamma\gamma}$ versus $\mu_{ggF,\gamma\gamma}$ for different values of
$\kappa_{b \bar b}$, for 120 GeV $< m_{\tilde{b}_1}<$ 300 GeV.
In the case of sbottoms, as the arrows indicate, their main effect is to 
reduce $\kappa_{gg}$, reducing $\mu_{ggF,\gamma\gamma}$, with a much smaller
effect on $\kappa_{\gamma\gamma}$, producing only a small increase in 
$\mu_{VBF,\gamma\gamma}$. The reduced Yukawa coupling to bottoms then reduces 
the total width, causing a universal increase in $\mu_{\gamma\gamma}$ 
irrespective of the production channel.
Overall, we see that, in this situation, light sbottoms can produce quite large non-universal
alterations, which can be measured by the
$\tfrac{\mu_{VBF,\gamma\gamma}}{\mu_{ggF,\gamma\gamma}}$ ratio
which can be even larger than in the light stop scenario.
Fig.~\ref{sbottom_2nd}(b)(ii) is similar to Fig.~\ref{sbottom_2nd}(b)(i), but with the 
colours indicating the sbottom mass in each case. We see that the largest effects
are produced by the lightest sbottoms, as expected, but that a significant effect 
giving $\tfrac{\mu_{VBF,\gamma\gamma}}{\mu_{ggF,\gamma\gamma}} \sim 1.2$ is still possible 
for sbottoms as heavy as $250$ GeV $\le m_{\tilde{b}_1} \le 300$ GeV.


%% file: 03.4_stau.tex
\subsection{Stau effects}

In addition to light stops and sbottoms, the lightest stau may give important
contributions that in particular could enhance $\kappa_{\gamma\gamma}$. 
For staus, $N_{c} Q^2 = 1$, a factor of 3 larger than sbottoms, and since
the Higgs-stau coupling like the Higgs-sbottom coupling also does not depend
on $X_t$, and hence is not constrained by the Higgs mass, light stau effects on
$\Gamma({h\rightarrow\gamma\gamma})$ could be more significant than sbottom effects, 
with the caveat of a different (running) bottom mass versus the tau mass.

The Higgs coupling to the lightest stau, normalised by $v/\sqrt{2} = M_W/g$,
with $v$ the SM Higgs VEV,
is given by%
\bea%
\hat{g}_{h\tilde{\tau}_1\tilde{\tau}_1} = 
\cos 2 \beta \left( -\frac{1}{2} \cos^2 \theta_{\tilde{\tau}} +  \sin^2 \theta_W \cos 2 \theta_{\tilde{\tau}} \right)  
+ \frac{m_{\tau}^2}{M_Z^2}
+ \frac{m_{\tau}X_{\tau}}{2 M_Z^2} \sin2 \theta_{\tilde{\tau}}.%
\eea %
with $X_{\tau} = A_{\tau} - \mu \tan \beta$.

Similarly to sbottoms, for a large positive $\mu$, with large $\tan \beta$,
$X_{\tau}$ is large and negative, and we find that
\begin{equation}
\hat{g}_{h\tilde{\tau}_1\tilde{\tau}_1} \simeq \frac{m_{\tau}X_{\tau}}{2 M_Z^2},
\end{equation}
which is large and negative.
Thus the stau contribution may enhance $\Gamma(h\to \gamma \gamma)$ in a large
$\tan \beta$ scenario with large and positive $\mu$.
(As in the sbottom case, a large negative $\mu$ would also give rise to a large
negative coupling, but we only consider positive $\mu$, as required such that $Y_b$ may be reduced).
As intimated, since $N_{c} Q^2$ is 3 times larger for the stau
than the sbottom, the minimum mass at which its contribution to $\kappa_{\gamma\gamma}$
can become large is approximately a factor of $\sqrt{3}$ times heavier than for the sbottom. 

\begin{figure}[htb]
\epsfig{file=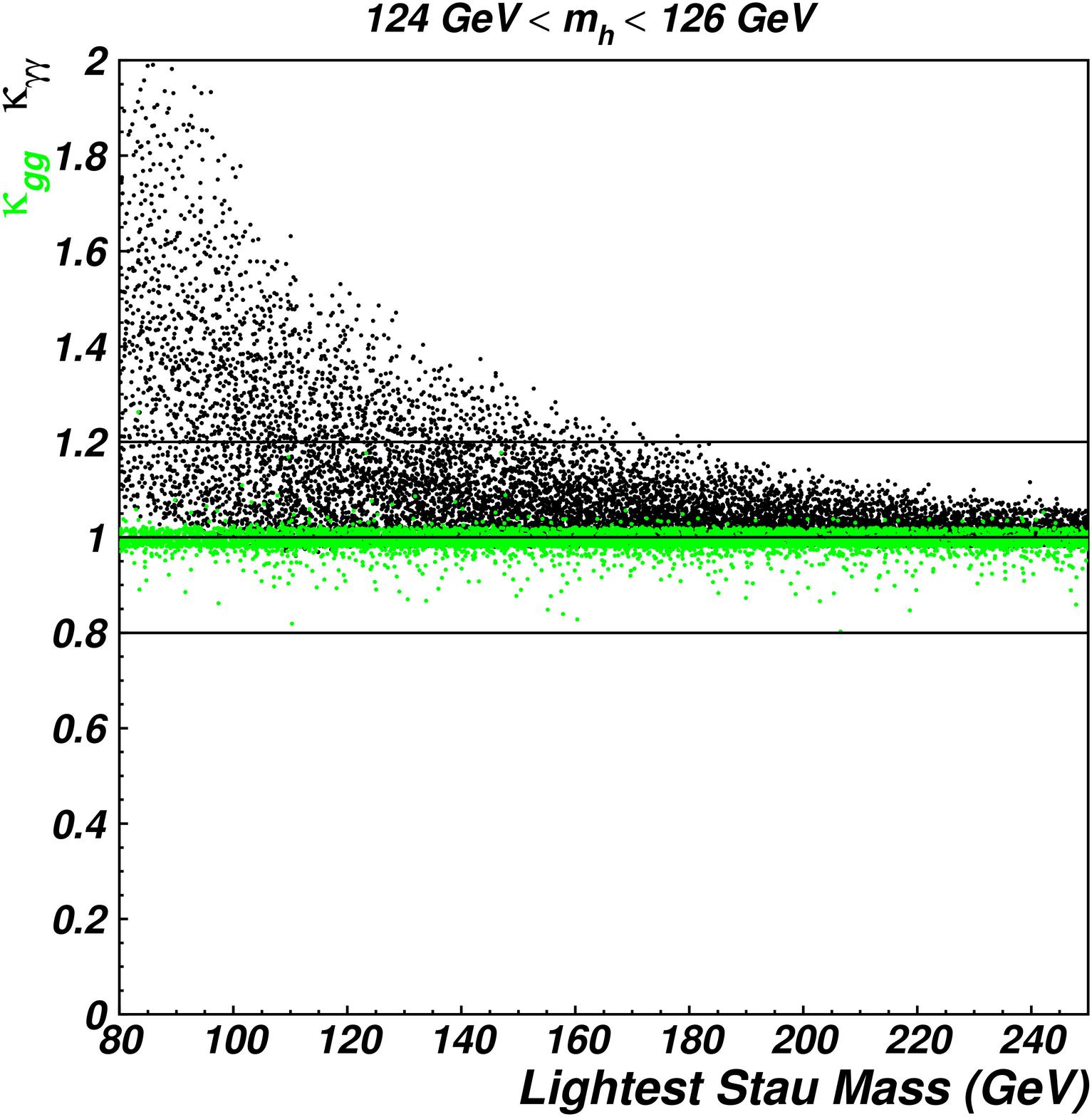,width=0.45\textwidth,angle=0}%
\epsfig{file=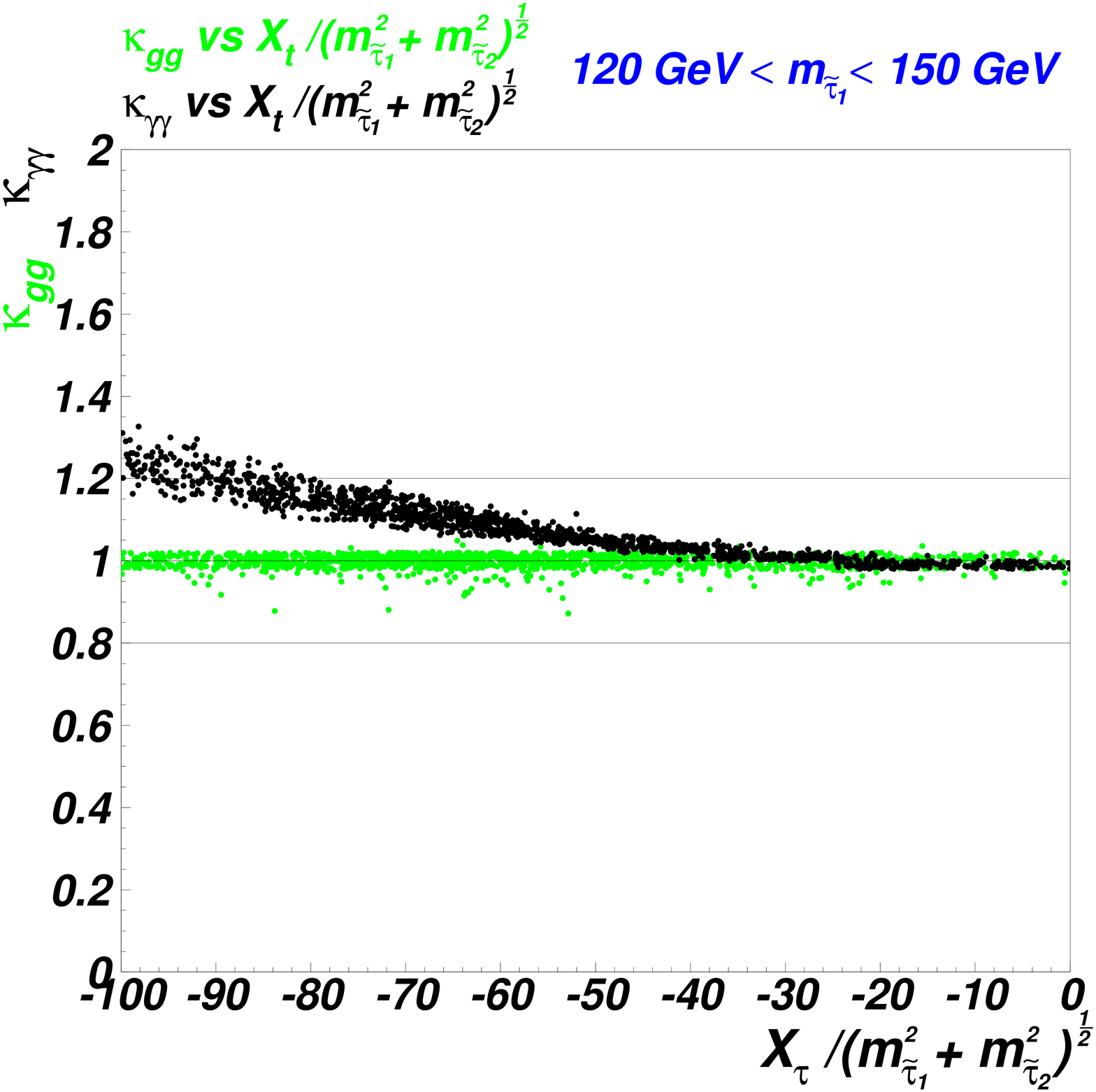,width=0.55\textwidth,angle=0}%
\vskip -0.3cm
\hspace*{0.1\textwidth}(a)\hspace*{0.3\textwidth}(b)
\vspace*{-0.4cm}
\caption{
(a) $\kappa_{\gamma\gamma}$ (black) and $\kappa_{gg}$ (green) as a function of lightest stau mass
for 124 GeV $< m_h <$ 126 GeV.
(b) $\kappa_{\gamma\gamma}$ (black) and $\kappa_{gg}$ (green) as functions of
$\frac{X_\tau}{(M_{\tilde{\tau}_1}^2 + M_{\tilde{\tau}_2}^2)}$ for $120$ GeV $\le m_{\tilde{t}_1} \le 140$ GeV.
To isolate the influence of light staus, the following cuts are also applied:
$m_{H^{\pm}}$, $m_{\chi^{\pm}_{1,2}}$, $m_{\tilde{t}_{1,2}}$, $m_{\tilde{b}_{1,2}}$, $m_{\tilde{\tau}_2} > 300$ GeV.
} %
\label{kAA-stau-first}
\end{figure}

This is
demonstrated in Fig. \ref{kAA-stau-first}(a), where $\kappa_{\gamma\gamma}$ (black) is plotted
against the stau mass, indeed showing that we can have $\kappa_{\gamma\gamma} > 1.2$
when $m_{\tau} \lesssim 180$ GeV. It also shows that light staus have no effect on
$\kappa_{gg}$ as expected. (The points with a slight reduction in $\kappa_{gg}$ have
sbottoms or stop masses $\sim$ 300 GeV, just above the mass cut applied for these particles).
In Fig. \ref{kAA-stau-first}(b), $\kappa_{\gamma\gamma}$ and $\kappa_{gg}$ are plotted
against $\frac{X_\tau}{\sqrt(M_{\tilde{\tau}_1}^2 + M_{\tilde{\tau}_2}^2)}$, showing that as 
for the sbottom, the coupling becomes largest for large and negative $X_\tau$.

\begin{figure}[htb]
\epsfig{file=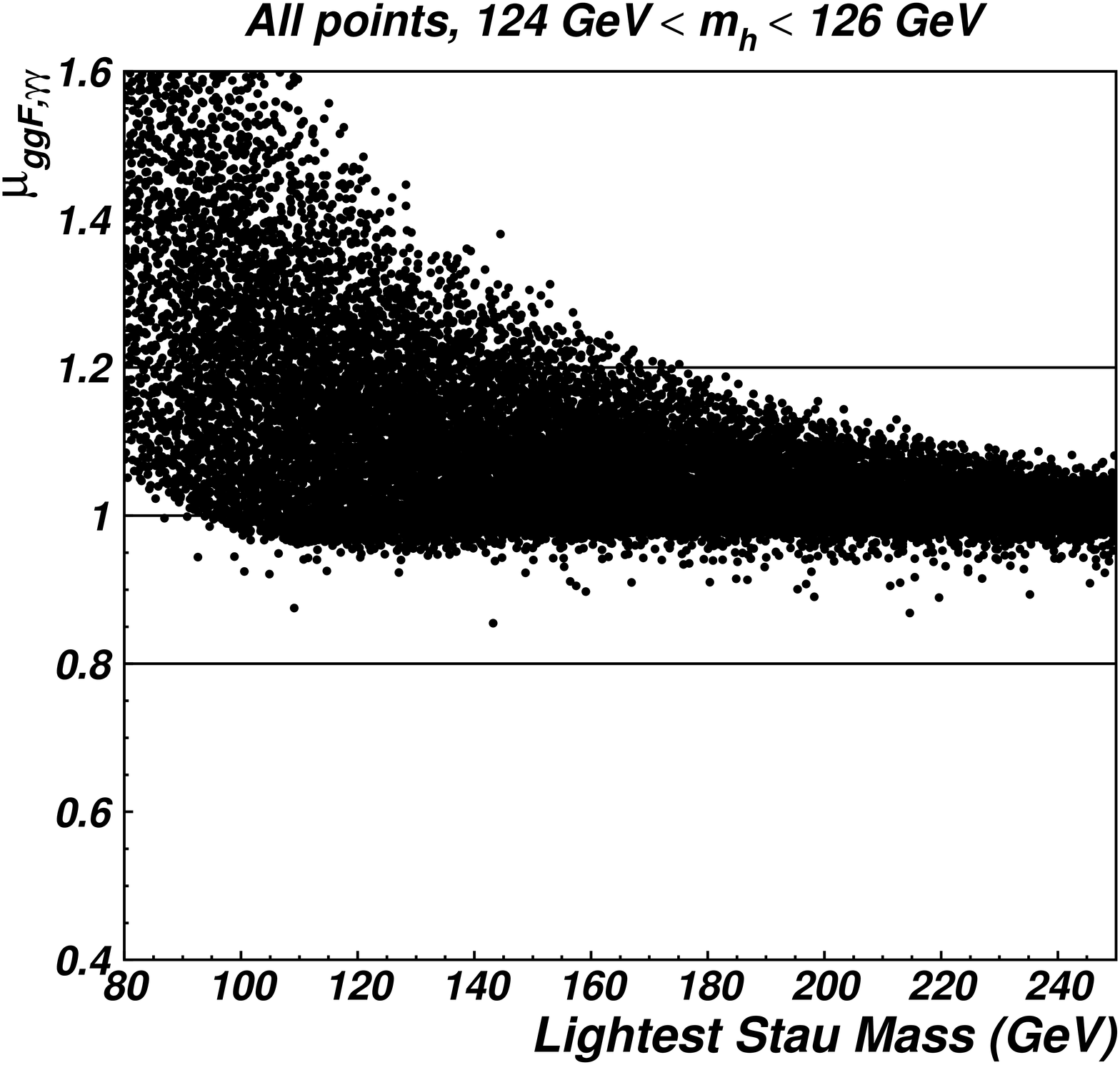,width=0.325\textwidth,angle=0}%
\epsfig{file=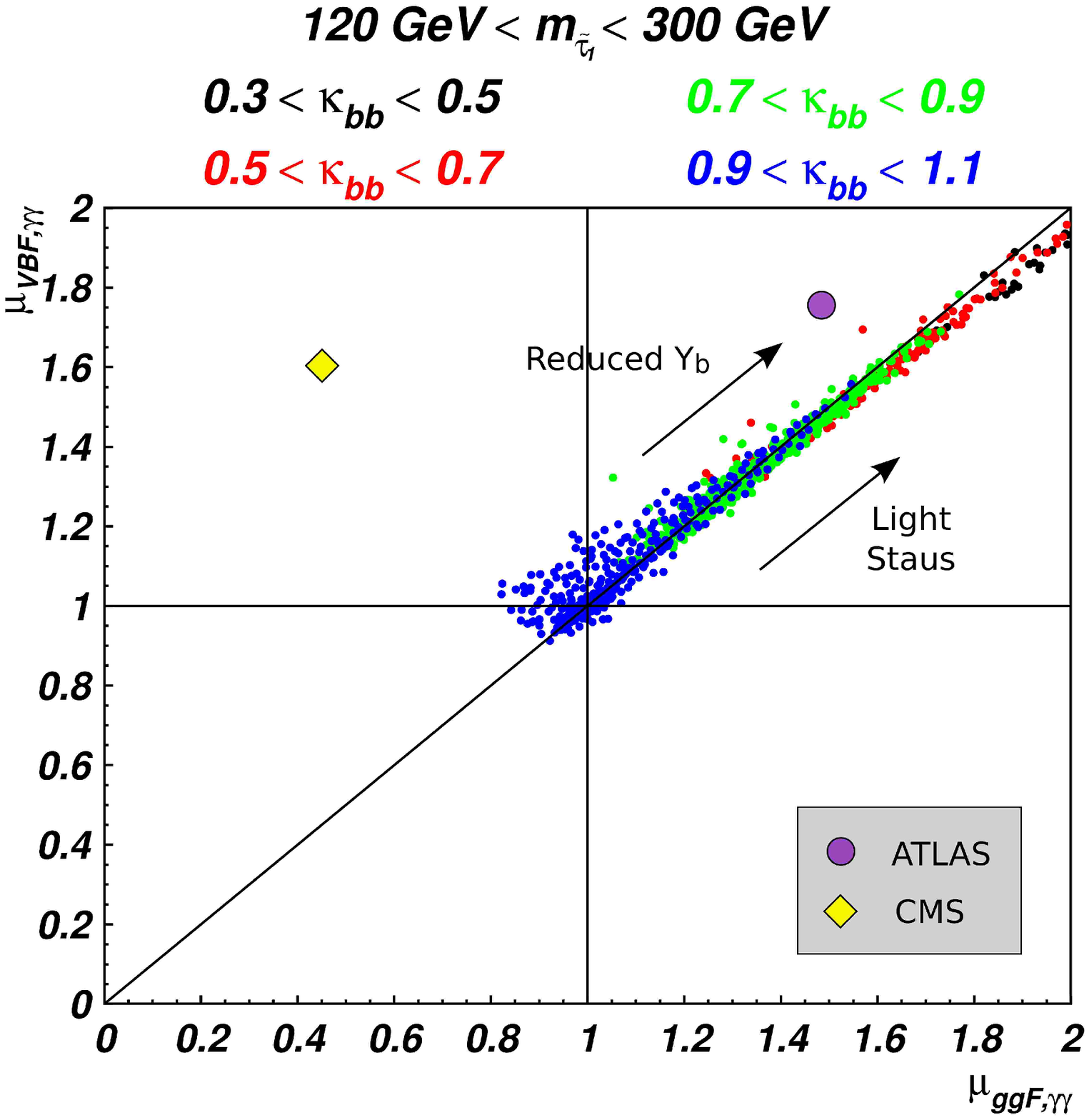,width=0.35\textwidth,angle=0}%
\epsfig{file=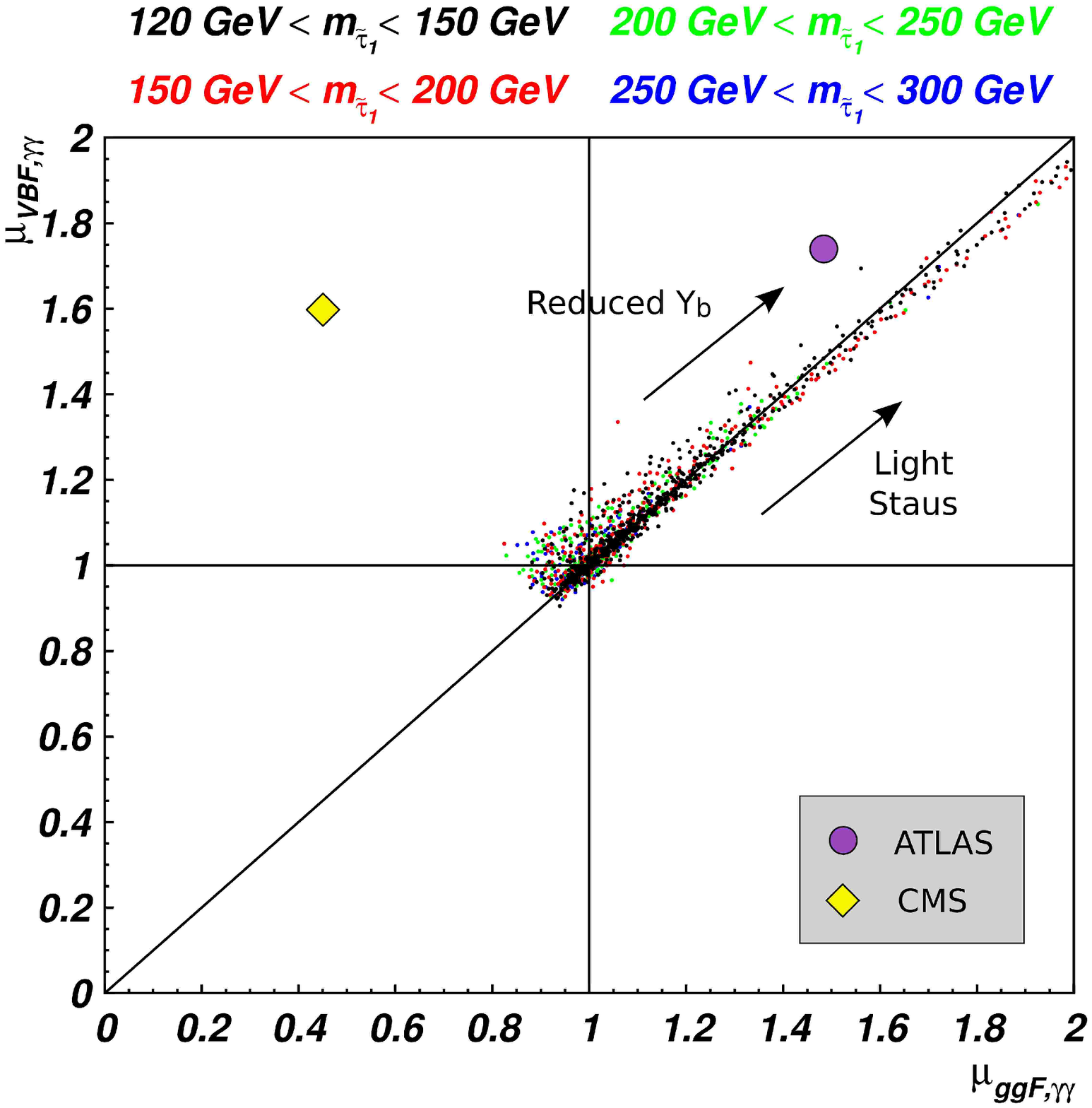,width=0.325\textwidth,angle=0}%
\vskip -0.3cm
\hspace*{0.1\textwidth}(a)\hspace*{0.25\textwidth}(b)\hspace*{0.25\textwidth}(c)
\\
\vspace*{-0.4cm}
\caption{
(a) $\mu_{ggF,\gamma \gamma}$ vs lightest stau mass.
We have cut for $0.98 \le \kappa_{bb} \le 1.02 $ to remove the possible effect of a reduced
$\Gamma_{hb\bar{b}}$.
(b) Each point of the scan with $120$ GeV $\le m_{\tilde{\tau}_1} \le 300$ GeV
is plotted on the ($\mu_{VBF},\mu_{ggF}$) plane, with colours to indicate
(i) different values for $\kappa_{bb}$,
(ii) different $m_{\tilde{\tau}_1}$ masses.
The results from ATLAS (purple circle) and CMS (yellow diamond) are indicated for comparison.
In all the plots, $124$ GeV $\le m_h \le 126$ GeV,
and to isolate the influence of light staus the following cuts are also applied:
$m_{H^{\pm}}$, $m_{\chi^{\pm}_{1,2}}$, $m_{\tilde{t}_{1,2}}$, $m_{\tilde{b}_{1,2}}$, $m_{\tilde{\tau}_2} > 300$ GeV.
} %
\label{kAA-stau-second}
\end{figure}
As staus are colourless and do not affect the gluon-gluon fusion production channel, 
$\mu_{ggF,\gamma\gamma}$ 
follows a very similar
pattern as  $\kappa_{\gamma\gamma}$. This is illustrated  in Fig.\ref{kAA-stau-second}(a) where 
we see that for $m_{\tilde{\tau}_1} \lesssim 180$ GeV, the value of 
$\mu_{ggF,\gamma\gamma}$ can become $> 1.2$.

In Fig.\ref{kAA-stau-second}(b(i,ii)), we see that the effect of the light staus on the 
($\mu_{VBF+VH},\mu_{ggF+ttH}$) plane is as expected, causing a universal increase in 
decay to di-photons irrespective of production channel, magnifying  also universal effects which may 
be caused by a reduction in $\Gamma_{hb\bar{b}}$.


%% file: 03.5_comb.tex
\subsection{Combined effect and fit of the LHC data}

\begin{figure}[htb]
\begin{center}
\centerline{\epsfig{file=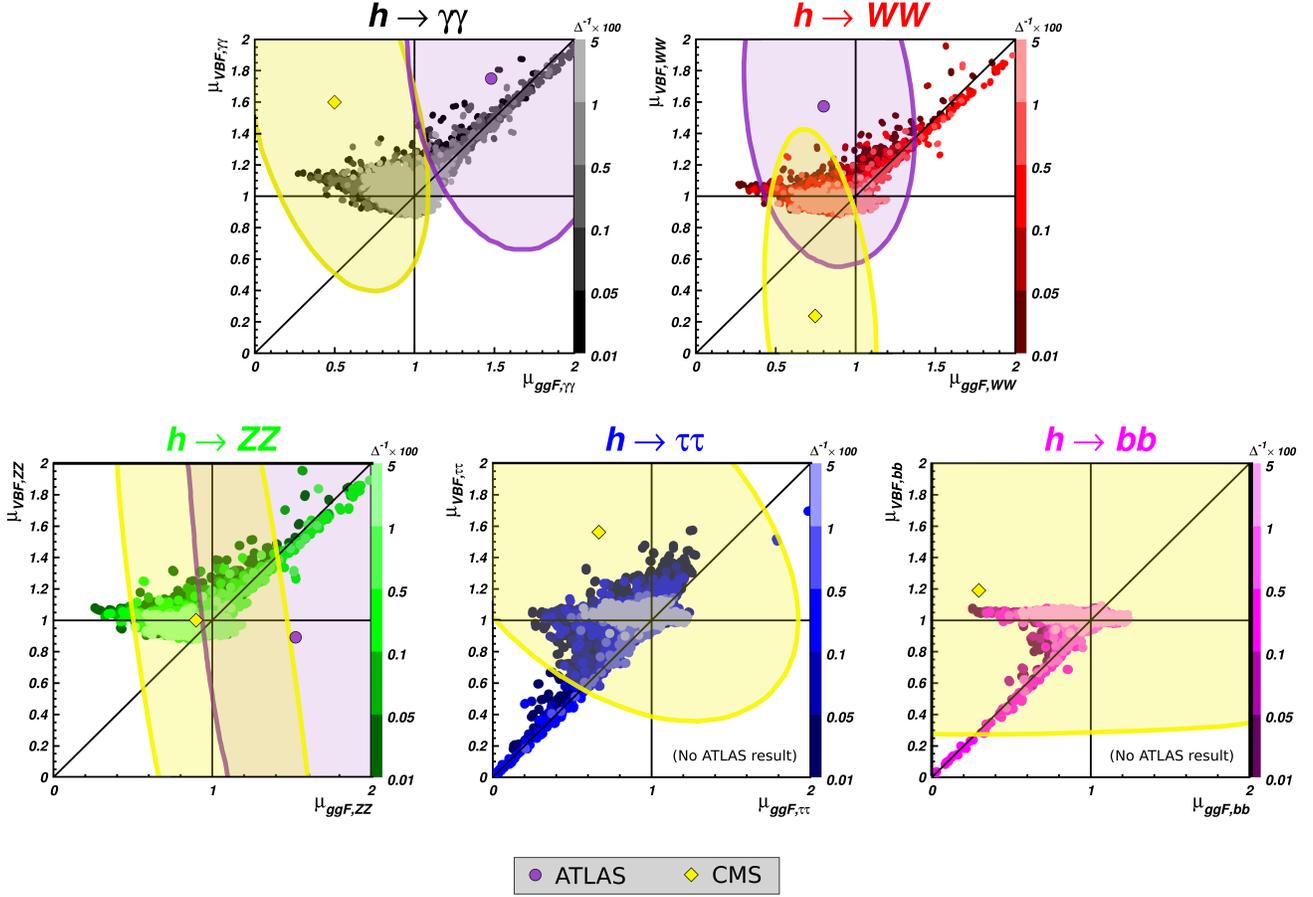,width=1.1\textwidth,angle=0}}%
\caption{$\mu_{VBF}$ vs $\mu_{ggF}$ for the di-photon, WW, ZZ, $\tau\tau$ and $b \bar b$
decay channels where the lightest stop and/or lightest stau and/or lightest sbottom 
has a mass between 120 GeV - 300 GeV.
ATLAS (circle) and CMS (square) best fit results for each channel
are also plotted. The 68\% Confidence Level (CL) for the experimental results are included.
ATLAS results were not available for the $b \bar b$ and $\tau\tau$ channels.
Colour gradients denote different values of the fine-tuning parameter
$\Delta^{-1} \times 100$ as described in subsection~\ref{sec:stops}.
}
\label{5_figures}
\end{center}
\end{figure}

In Fig.~\ref{5_figures} we present results for $\mu_{VBF}$ versus $\mu_{ggF}$ 
for the $\gamma\gamma$, $WW$, $ZZ$, $\tau\tau$ and $b \bar b$ decay channels
in the ($\mu_{VBF},\mu_{ggF}$) plane, where we have included all points from our scan 
for which any or all of the scenarios discussed in the previous Subsections are realised.
We see that, for each final state, the majority of parameter space has
$\mu_{VBF} > \mu_{ggF}$, and comparing with experimental measurements, we see that
6 out of 8 measurements
have $\mu_{VBF} > \mu_{ggF}$ for their best fit values.
Therefore one can expect that the MSSM will provide a better fit to the data and,
in general, that the  light stop and sbottom scenarios
would be able to explain a non-universal alteration of $\mu_{VBF} > \mu_{ggF}$,
if this is confirmed at the upgraded LHC starting in 2015.
In Fig.~\ref{5_figures} we have also stratified by shading according to the
values of the fine-tuning parameter $\Delta$, as described in section~\ref{sec:stops}.
We see that the points with a large universal increase in $\mu_{ggF}$ and $\mu_{VBF}$ have a larger
fine tuning in general. This is expected, as for these points $\mu$, on which $\Delta$ 
depends, is required to be large to reduce $Y_b$ as discussed in section~\ref{sec:MSSM_effects}.

To quantify how well the scenarios we have discussed fit current LHC data, we have calculated
the $\chi^2$ for each scenario and compared to the best fit values for $\mu_{ggF}$ and $\mu_{VBF}$ 
in Fig.~\ref{fig:lhc-comb}. For each collaboration (ATLAS, CMS) the systematic errors
on the values of $\mu_{ggF}$ and $\mu_{VBF}$ for a single decay mode are correlated. To 
take this correlation into account, we actually calculate a ``profiled log likelyhood ratio''
test statistics. However, under the assumption that the 
data is distributed as a multivariate Gaussian (valid to a good approximation),
this reduces to a $\chi^2$ statistics 
with a non-diagonal covariance matrix. This is discussed fully in \cite{Belyaev:2013ida}.
There are 6 degrees of freedom from the ATLAS
data and 10 from the CMS data, giving 16 degrees of freedom overall (3 channels
for ATLAS ($\gamma\gamma$, $WW$ and $ZZ$) and 5 from CMS ($\gamma\gamma$, $WW$, $ZZ$, $\tau\tau$, $b\bar b$), 
each with $ggF$ and $VBF$ channels). 
The regions of interest for which a $\chi^2$ was calculated 
were defined as; 1) light stops only, 2) light sbottoms only, 3) light
staus only, 4) lights stops and/or sbottoms and/or staus. In each case a ``light'' mass was
defined as between 120 and 300 GeV.

\begin{figure}[htb]
\begin{center}
\epsfig{file=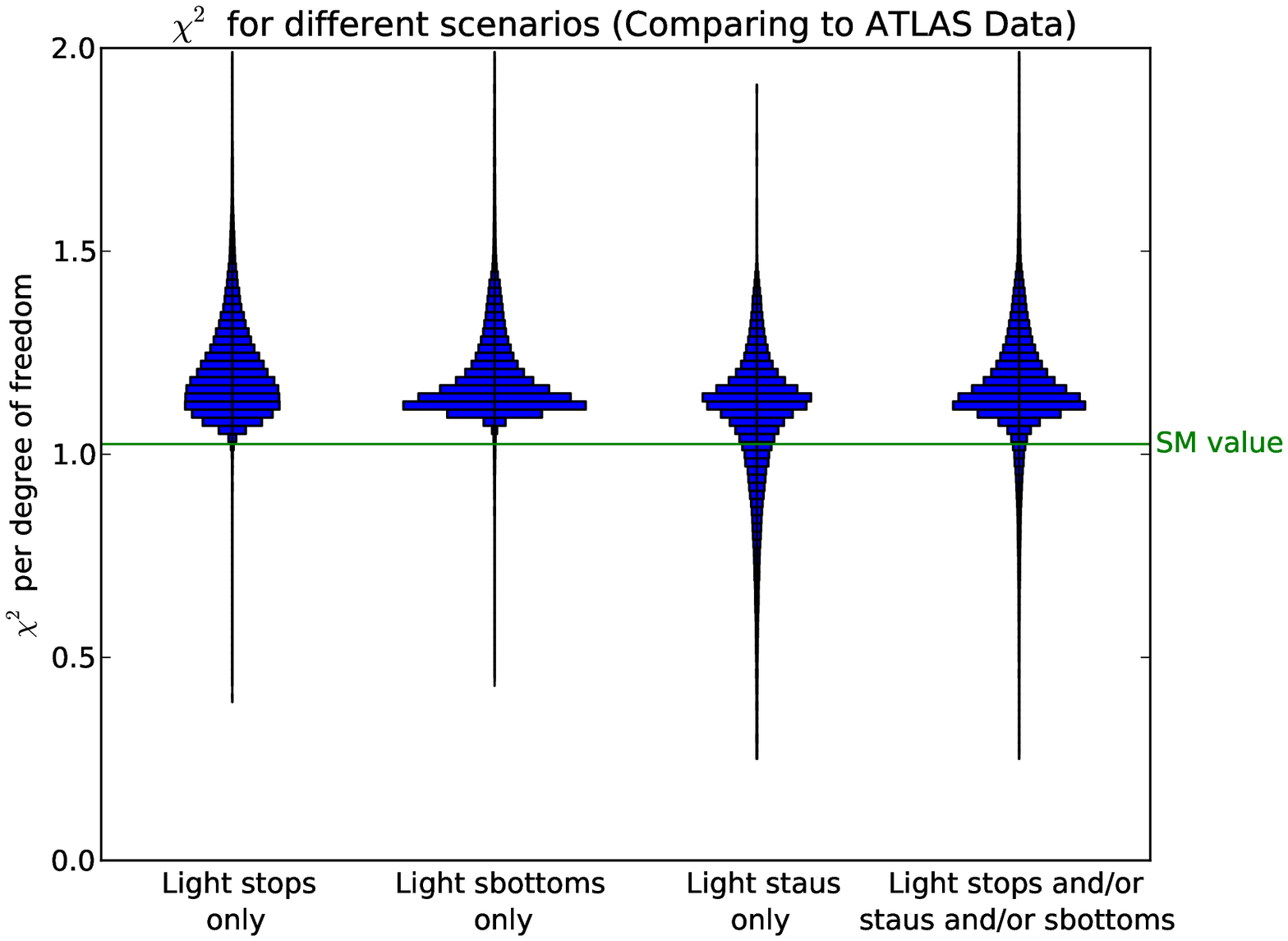,width=0.5\textwidth,angle=0}%
\epsfig{file=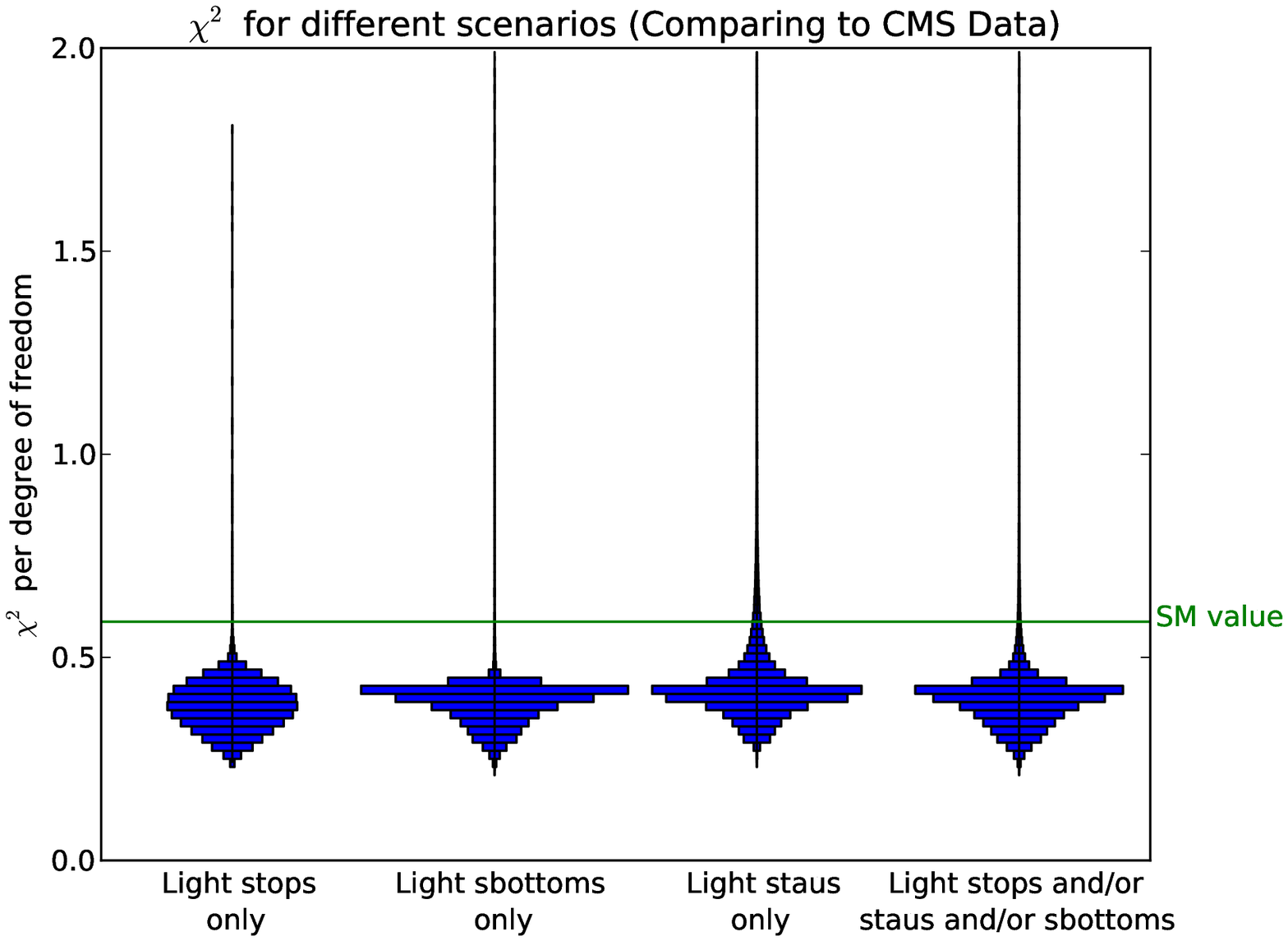,width=0.5\textwidth,angle=0}\\
\vskip -0.3cm
\hspace*{0.1\textwidth}(a)\hspace*{0.3\textwidth}(b)
\epsfig{file=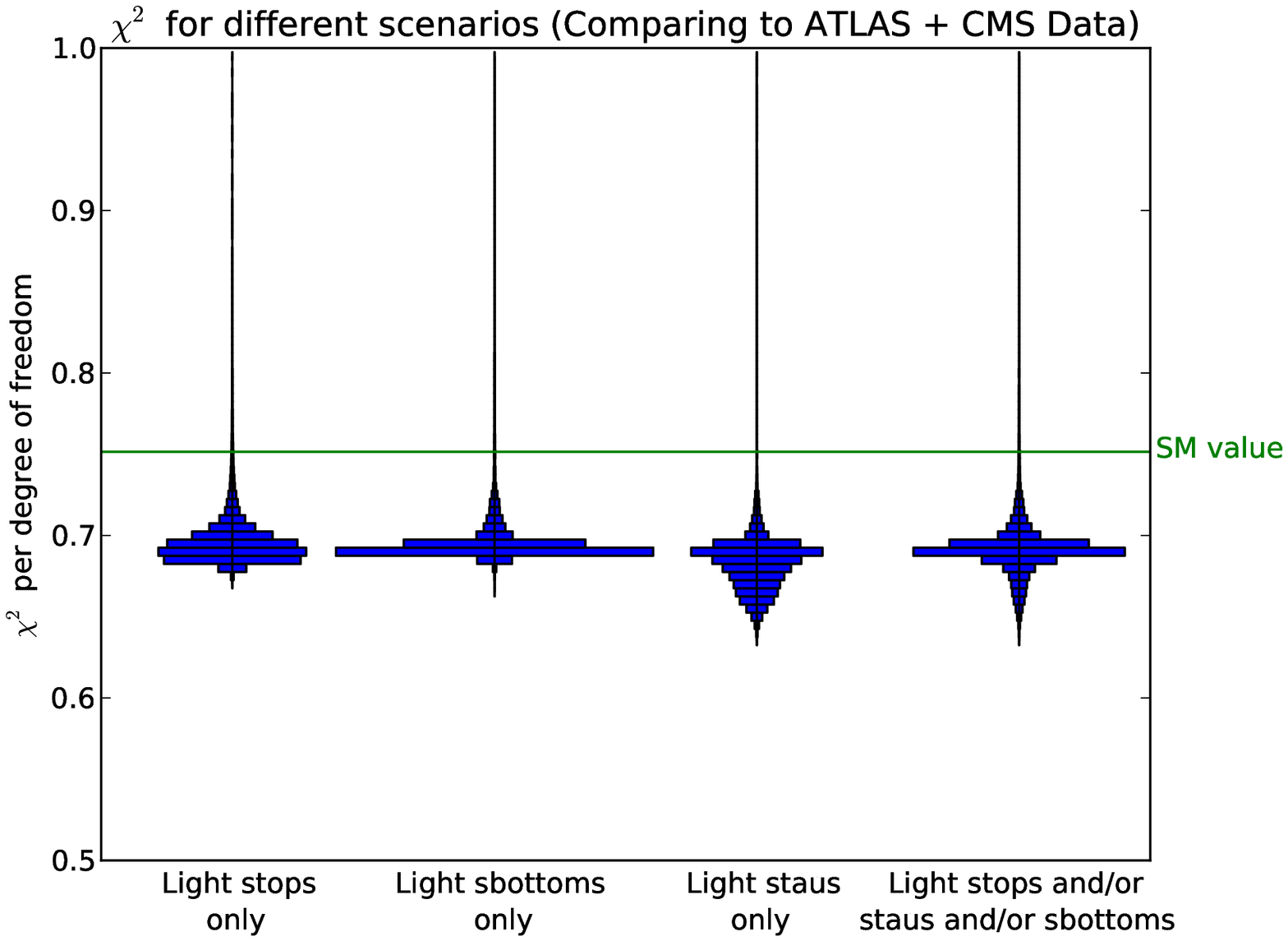,width=0.8\textwidth,angle=0}%
\vskip -0.3cm
\hspace*{0.2\textwidth}(c)
\caption{$\chi^2$ results per degree of freedom for
different regions in parameter space compared to (a) ATLAS data,
(b) CMS data, (c) combined ATLAS+CMS data. The width of each block is
proportional to the number of points in each region of parameter space with
each particular value of $\chi^2$. The light stop, sbottom and stau regions
are defined as $120$ GeV $\le m_i \le 300$ GeV, where
$m_i = m_{\tilde{t}_1}$, $m_{\tilde{b}_1}$, $m_{\tilde{\tau}_1}$ respectively.
The SM fit to data is indicated by the green line for each plot.
} %
\label{chi_blocks}
\end{center}
\end{figure}

For each case, we have calculated the $\chi^2$ for each point in the parameter space from our scan
that matched the aforementioned particle mass criteria. The results are presented in Fig.~\ref{chi_blocks}
as the $\chi^2$ per degree of freedom ($\chi^2/N_{\rm DOF}$), when compared to (a) just the
ATLAS results, (b) just the CMS results, (c) both ATLAS and CMS results.

First of all, one can see that the SM already fits the data well, 
the $\chi^2/N_{\rm DOF}$ is about 1 for 
ATLAS (Fig.~\ref{chi_blocks}(a)) and is about 0.6 for CMS (Fig.~\ref{chi_blocks}(b)),
while for the ATLAS+CMS combination we have got  $\chi^2/N_{\rm DOF} \sim 0.75$
(Fig.~\ref{chi_blocks}(c)).
One should note that the $\chi^2/N_{\rm DOF}$ is lower for CMS, mainly due to the 
best fit point for $h\to ZZ$ being very close to the SM value,
and for $h\to b \bar b$ being reasonably close to the SM with very large error bars.
It is hard to judge which scenario is preferred due to the large experimental errors,
but one can see that the combined scenario as well as the scenario with light staus give the
lowest value of  $\chi^2/N_{\rm DOF}$.

%% file: 04_conclusions.tex
\section{Conclusions}
In this paper we have explored the 
effect from light sfermions on the production and 
decay of the lightest Higgs boson within the MSSM culminating with a fit of 
the relevant parameter space to LHC data.

We have found that the scenario with light coloured sfermions, namely stops and sbottoms,
has the potential to explain a non-universal alteration
of the two most relevant Higgs production channels, {\it  i.e.}, $\mu_{VBF} \ne \mu_{ggF}$, 
and predicts $\frac{\mu_{VBF}}{\mu_{ggF}}>1$ in all Higgs boson decay channels
in the majority of the parameter space. These light stop and light sbottom scenarios are 
realised in specific regions of the parameter 
space in terms of soft-breaking mass terms, namely where
$M_{Q3} >> M_{U3}$ and $M_{Q3} >> M_{D3}$, respectively.

The specific feature of the  scenario with a light stop is that  
$\hat{g}_{h\tilde{t}_1\tilde{t}_1}$ is  negative  (whenever one is near maximal stop mixing),  
which makes  the overall stop loop
contribution of the opposite sign as   the top quark contribution and of 
 same sign as the $W$ loop. As a result, one obtains a decreased $k_{gg}$
and an  increased $k_{\gamma\gamma}$ couplings. It is important to notice that the 
relative decrease of $k_{gg}$   is bigger than the relative increase
of $k_{\gamma\gamma}$, so the overall effect from  light stops per se would lead to 
a decrease of the Higgs production via gluon-gluon fusion decaying to 
di-photons ($\mu_{ggF,\gamma\gamma}$) as well as a reduction in $\mu_{ggF}$ 
compared to the SM for all decay channels ($\mu_{ggF}^{SM} = 1$  by definition). 
This scenario would be
somewhat consistent with CMS data, where $\mu_{ggF} < 1$ for all decay channels.

However, this prediction is in tension with the ATLAS data where $\mu_{ggF} \approx 1.5$
for both the $\gamma\gamma$ and $ZZ$ decay channels. 
Therefore, we also consider the scenario where we have both light stops and
a suppressed $hb\bar b$ coupling. In this case, the reduced $h\to b\bar{b}$ partial
width (and related $\kappa_{bb}$ parameter) causes an enhancement of the
BRs and hence signal strengths of the other decay channels, which
can compensate for the reduced production via $ggF$. In this case, depending
on the degree of suppression of the $hb\bar b$ coupling, the $ggF$ signal strength in all 
channels can be increased, either to match the SM level, or greater, {\it e.g.}, to 
$\mu_{ggF} \approx 1.5$, in order to be more consistent with the ATLAS data. The exceptions
being $\mu_{ggF,\gamma\gamma}$, which can be slightly enhanced, and $\mu_{ggF,bb}$, which 
would be reduced compared to the other channels. This reduction of the $hb\bar b$
coupling was achieved with a large $\mu$ (1--5 TeV), intermediate $M_A$ (300--800 GeV) and
intermediate-to-large $\tan\beta$ (20--50).

One should also note that, in the light stop scenario,
 $\hat{g}_{h\tilde{t}_1\tilde{t}_1}$  is fixed (in the  maximal mixing scenario) 
to about $-\frac{1}{2}\frac{m_t^2}{M_Z^2}$, which limits the maximal
contribution of the stop to production via $ggF$ and to the $h \rightarrow \gamma\gamma$ decay.
We have also found that the effect from  light sbottoms on the gluon fusion rate
can be potentially larger (since Higgs-sbottom-sbottom coupling is not correlated with the Higgs boson mass), 
which in its turn requires a bigger decrease 
for $\kappa_{bb}$ to satisfy the experimental data.
One can see that, in the light stop/sbottom scenarios,
${\mu_{ggF,bb}}$ and  ${\mu_{VBF,bb}}$ are predicted to be   essentially below
unity,  especially the  ${\mu_{ggF,bb}}$ value, which is doubly suppressed both via $ggF$ production
(due to the negative interference from stop/sbottom loops) and from decay
(due to  $\kappa_{bb}$ suppression). 
Therefore, in the future LHC runs, the measurement of ${\mu_{ggF,bb}}$ and ${\mu_{VBF,bb}}$
is particularly important, as this will enable one to exploit an additional constraint in order to pin down possible MSSM
effects in the Higgs sector.

In contrast, light staus were found to be able to only universally 
increase the signal strengths, irrespective of the production channel, generally complementing the 
effect of a reduced $hb\bar b$ coupling.

Furthermore, we showed that the non-universal solutions (${\mu_{ggF}} \ne {\mu_{VBF}}$)
had a fairly low minimum fine-tuning measure, as low as $\sim 5\%$, while 
in regions where a universal increase in signal strength (${\mu_{ggF}} \sim {\mu_{VBF}} > 1$) is 
caused due to a suppressed $Y_b$, the fine-tuning was much 
larger due to the requirement of a large $\mu$ parameter.

Finally, we performed a $\chi^2$ fit for these MSSM scenarios which showed that they 
all fit data better than the SM.

To conclude, we have found that the MSSM with light stops or sbottoms
has a good potential to explain a non-universal alteration 
of the Higgs production rates in different channels as compared to the SM
and that this can be complemented by the fact that a reduced  $\Gamma_h$ could 
increase the relevant signal strengths to be equal to or greater
than the SM prediction. 

As an outlook, we should mention the fact that, among the light squark solutions considered, the light stop one 
is also attractive from a cosmological point of view. In fact,
the scenario where 
the neutralino (the LSP) is degenerate with the lightest stop
in the 100--300 GeV range (to satisfy the experimental data on stop quark searches)
predicts a plausibly low amount of DM (via the stop-neutralino co-annihilation channel).
At the same time the light stop scenario 
can provide a crucial link to EWBG, specifically in case of very light stops.
In this connection, we also suggest an alternative solution of compensating the
enhancement of $ggF$ production due to light stops (which takes place away from maximal mixing scenario), through
an increase of $Y_b$ and a  respective decrease of the $\gamma\gamma/ZZ/WW/\tau\tau$  decay rates.